\documentclass{emulateapj}
\usepackage{float,longtable,amsmath}

\shorttitle{BAT Catalog and Correlations}
\shortauthors{Butler et al.}

\def\gtrsim{\mathrel{\hbox{\rlap{\hbox{\lower4pt\hbox{$\sim$}}}\hbox{$>$}}}}
\def\lessim{\mathrel{\hbox{\rlap{\hbox{\lower4pt\hbox{$\sim$}}}\hbox{$<$}}}}

\newcommand\swift{{\it Swift}}

\slugcomment{Accepted to ApJ}

\begin{document}

\title{A Complete Catalog of Swift GRB Spectra and Durations: \\
Demise of a Physical Origin for Pre-Swift High-Energy Correlations}

\author{Nathaniel R. Butler\altaffilmark{1,2},
Daniel Kocevski\altaffilmark{2}, Joshua S. Bloom\altaffilmark{2,3}, and Jason L. Curtis\altaffilmark{2}.}
\altaffiltext{1}{Townes Fellow, Space Sciences Laboratory,
University of California, Berkeley, CA, 94720-7450, USA}
\altaffiltext{2}{Astronomy Department, University of California,
445 Campbell Hall, Berkeley, CA 94720-3411, USA}
\altaffiltext{3}{Sloan Research Fellow}

\begin{abstract}
We calculate durations and spectral parameters for 218 Swift bursts detected
by the BAT instrument between and including GRBs~041220 and 070509,
including 77 events with measured redshifts.  Incorporating prior knowledge
into the spectral fits, we are able to measure the characteristic $\nu F_{\nu}$
spectral peak energy $E_{\rm pk,obs}$ and the isotropic equivalent energy
$E_{\rm iso}$
(1--$10^4$ keV) for all events.  
This complete and rather extensive catalog, analyzed with a  
unified methodology, allows us to address the persistence and origin  
of high-energy correlations suggested in pre-Swift observations.  We find
that the $E_{\rm pk,obs}$--$E_{\rm iso}$ correlation is present in the Swift
sample; however, the best-fit powerlaw relation is inconsistent with the
best-fit pre-Swift relation at $>5\sigma$ significance.  It has
a factor $\gtrsim 2$ larger intrinsic scatter, after accounting for large errors on $E_{\rm pk,obs}$.   A large fraction of
the Swift events are hard and subluminous relative to (and inconsistent with) the pre-Swift 
relation, in agreement with indications from {\it BATSE}~GRBs without redshift.
Moreover, we determine an experimental threshold for the BAT detector and show
how the $E_{\rm pk,obs}$--$E_{\rm iso}$ correlation arises artificially
due to partial correlation with the threshold.  We show that pre-{\it Swift}~correlations
found by \citet[]{amati02,yon04,firmani06} (and independently by others) are likely
unrelated to the physical properties of GRBs and are likely useless for tests of cosmology.
Also, an explanation of these correlations in terms of a detector threshold provides a 
natural and quantitative explanation
for why short-duration GRBs and events at low redshift tend to be outliers to the correlations.
\end{abstract}


\maketitle

\section{Introduction}

The {\it Swift}~satellite \citep{gehrels04} is revolutionizing our understanding of
Gamma-ray Bursts (GRBs) and their afterglows.  Our knowledge of the early X-ray afterglows
has increased tremendously due to the dramatic success of the X-ray Telescope \citep{burrows05}. However,
our understanding of the prompt emission properties has lagged.  This is due in part
to the narrow energy bandpass of the Burst Alert Telescope \citep[BAT;][]{bart05}, which precludes direct
measurement of the broad GRB spectra and tends to weaken any inferences about the $\nu F_{\nu}$ spectral
peak energy $E_{\rm pk,obs}$ and the bolometric GRB fluence.  Pre-{\it Swift}~observations
and estimations of these parameters lead to tantalizing correlations between the host-frame
characteristics of GRBs \citep[e.g.,][]{lpm00,fr00,nmb00,shaf03,amati02,lamb04,ggl04,firmani06}.  

The number of redshifts available in the {\it Swift}~sample now exceeds by
large factor the number of pre-{\it Swift}~GRBs with measured redshifts, and a {\it Swift}~BAT
catalog is a veritable gold-mine for the study of GRB intrinsic properties and possibly cosmological
parameters, provided we find a way to accurately constrain the BAT GRB energetics.

\citet{cabrera07} derive $E_{\rm pk,obs}$ values and isotropic equivalent energies $E_{\rm iso}$
for 28 BAT GRBs with measured redshift, in an impressive study which carefully accounts for
the narrow BAT bandpass.  Interestingly, their fits suggest an inconsistency between 
an $E_{\rm pk,obs}$-$E_{\rm iso}$ correlation in the {\it Swift}~sample relative to the pre-{\it Swift}~sample
\citep[e.g.,][]{amati02,lamb04}.
Several {\it Swift}~events appear to populate a region of the $E_{\rm pk,obs}$-$E_{\rm iso}$ plane
containing events harder and less energetic than those found prior to {\it Swift}.  Indications
that this might happen were found in the {\it BATSE}~GRB sample by \citet{np05} and \citet{bp05}. \citet{bp05}
estimate that as many as 88\% of {\it BATSE}~GRBs are inconsistent with the (pre-{\it Swift}) 
$E_{\rm pk,obs}$-$E_{\rm iso}$ relation and that this relation may in fact be an inequality, provided
we account for truncation by the detector threshold.

Below, we show that $E_{\rm pk,obs}$
determinations well above the nominal BAT upper energy of 150 keV, which agree well with those made by detectors actually sensitive at those energies,
are possible.
We constrain $E_{\rm pk,obs}$ and the $1-10^4$ keV fluence $S_{\rm bol}$ for
218 BAT GRBs, including 77 GRBs with host galaxy/spectroscopic redshifts.  As we describe in Section \ref{sec:spec}, this can be done because
the {\it BATSE}~catalog sets strong priors for the possible values of $E_{\rm pk,obs}$.
Moreover, we show (Section \ref{sec:amati}) that it is possible to rigorously account for the measurement uncertainty in $E_{\rm pk,obs}$
and $E_{\rm iso}$ when fitting for an ensemble relation between these quantities.

We find that a powerlaw
relation between $E_{\rm pk,obs}$ and $E_{\rm iso}$ is likely present but there is a large {\it intrinsic}~scatter --- even after accounting for the observed
scatter arising from the BAT narrow bandpass and resulting large $E_{\rm pk,obs}$ uncertainties.  The 
{\it Swift}~sample relation is inconsistent with all pre-{\it Swift}~relations
at the $>5\sigma$ level.
We experimentally infer the threshold of the detector (Section \ref{sec:thresh})
and test for the first time with many events the way the threshold impacts the observable host frame
quantities $E_{\rm pk}$ and $E_{\rm iso}$.  We find that the $E_{\rm pk}$--$E_{\rm iso}$ correlation, as well as the correlations found by
\citet{firmani06}, \citet{yon04}, and \citet{atteia03} are significant but simply due to a \citet{malm22} type bias in the source frame luminosity.

\section{Data Reduction and Temporal Region Definition}

Our automated pipeline at Berkeley is used to download the \swift~data in near real
time from the {\it Swift}~Archive\footnote{ftp://legacy.gsfc.nasa.gov/swift/data}
and quicklook site.  
We use the calibration files from the 2006-10-14 BAT database release.
We establish the energy scale and mask weighting for the BAT event mode data 
by running the {\tt bateconvert} and {\tt batmaskwtevt} tasks
from the HEASoft 6.0.6 software release\footnote{http://swift.gsfc.nasa.gov/docs/software/lheasoft/download.html}.  
Spectra and light curves are extracted with the {\tt batbinevt} task, and response
matrices are produced by running {\tt batdrmgen}.  We apply the systematic
error corrections to the low-energy BAT spectral data as suggested by the BAT
Digest website\footnote{http://swift.gsfc.nasa.gov/docs/swift/analysis/bat\_digest.html}, and fit the data in
the 15--150 keV band using 
ISIS\footnote{http://space.mit.edu/CXC/ISIS}.  The spectral normalizations are corrected for satellite slews using the {\tt batupdatephakw} task.  
All errors regions reported correspond to the 90\% confidence interval.  In determining source
frame flux values, we assume a cosmology with $h=0.71$, $\Omega_m=0.3$, and $\Omega_{\Lambda}=0.7$.

The timing and spectral analyses described in detail below first require the definition of a time region
encompassing the burst.

\subsection{Automated Burst Interval Determination}

The observed raw counts detected by the BAT are modulated by the coded-mask pattern
above the detector.  By ``mask-weighting'' the observed data using a known source position,
assumed here to be the position from the XRT if available,
the standard analysis software effectively removes the mean counts flux from 
background sources.  Estimation of the burst time interval and count rate from the 
mask-weighted light curve therefore does not require the fitting of a background term.  Because
the burst interval is defined only by a start time t1 and a stop time t2, it is possible
to quickly measure the signal-to-noise ($S/N$) ratio of every possible burst interval
for a given stretch of data known to contain a GRB.  

\begin{figure}
\centerline{\rotatebox{270}{\includegraphics[width=2.6in]{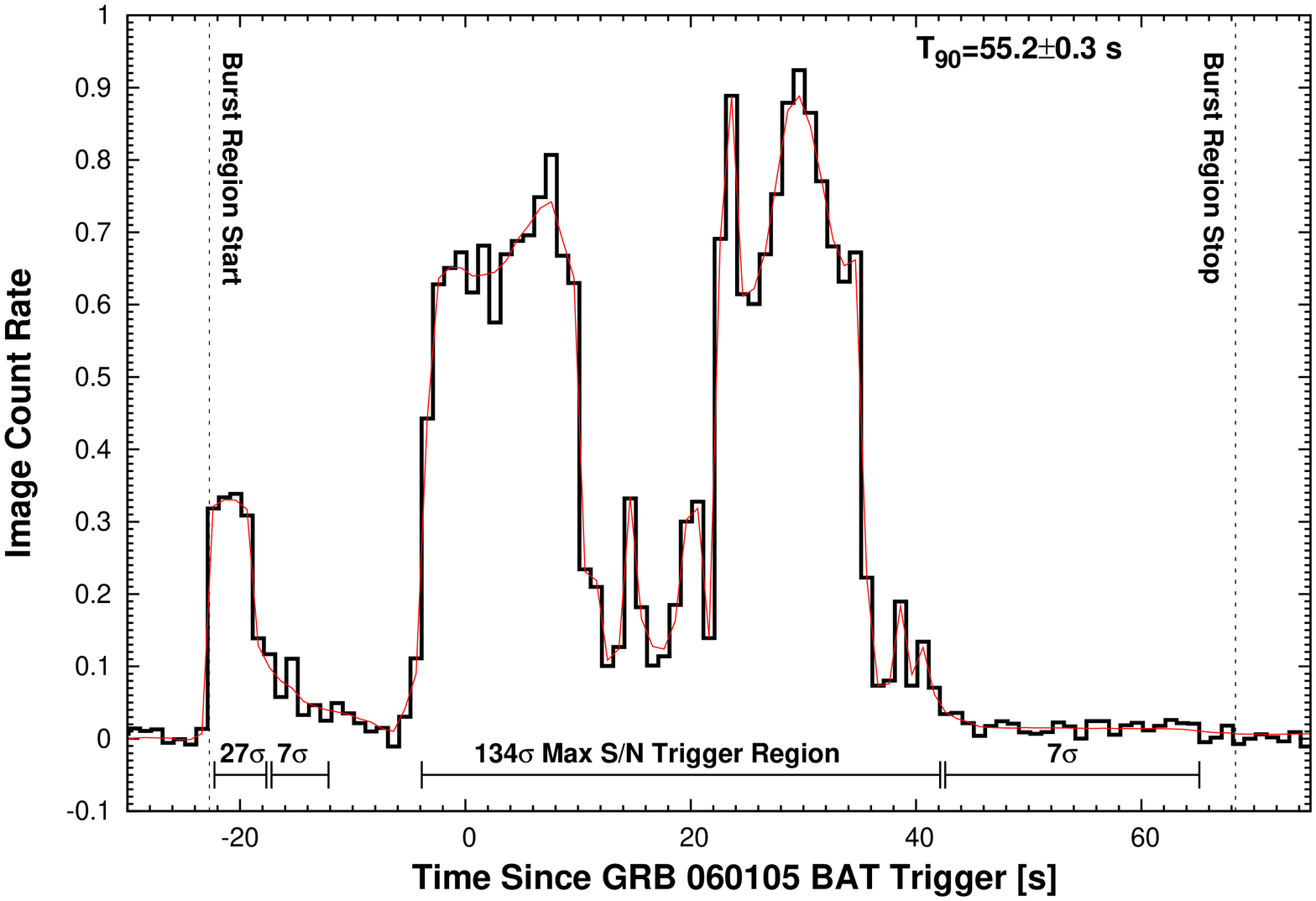}}}
\centerline{\rotatebox{270}{\includegraphics[width=2.6in]{f1b.ps}}}
\centerline{\rotatebox{270}{\includegraphics[width=2.6in]{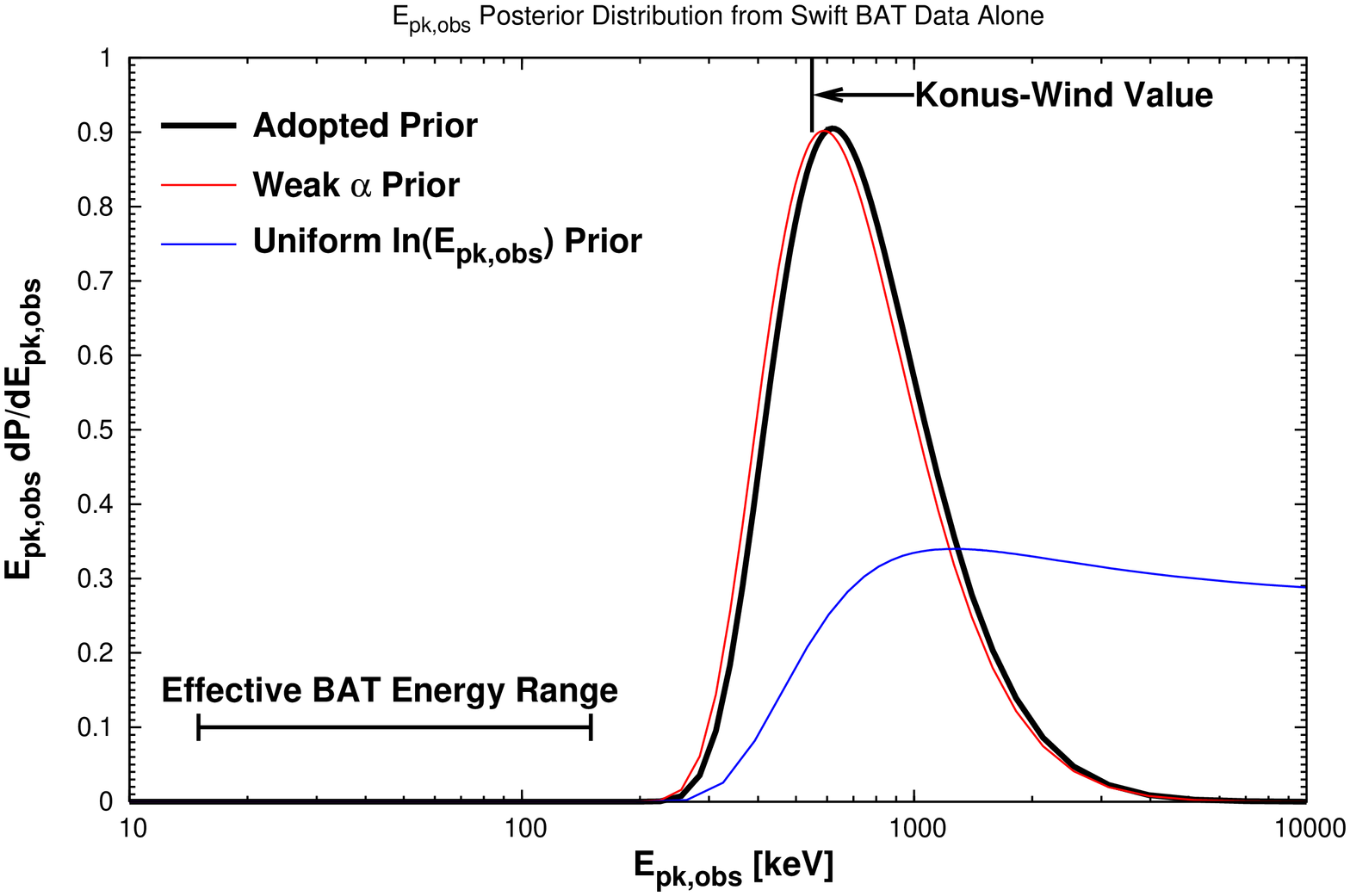}}}
\caption{\small Time region definition, duration estimation, and spectral parameter
estimation for example GRB~060105. (Top Panel) Light curve plotted with 1s binning.
The automated algorithm finds four trigger regions over
threshold and merges these into a burst window of width $\Delta t=87.4$s.  This
region is then extended backward by 0.46 s and forward by 3.22 s based on signal
in the denoised light curve (overplotted in red) to define the final burst interval.
(Middle Panel) The time-integrated spectrum is acceptably fit by a simple powerlaw
(Table 2).  (Bottom Panel) Using prior information (primarily on $E_{\rm pk,obs}$)
we are able to infer the presence of a spectral break located well above the BAT
detector bandpass, consistent with measurements by Konus-Wind \citep{gol06}.}
\label{fig:triggerNprob}
\end{figure}

We employ the following automated 
3-step procedure
to define an optimum burst interval in the sense that it is likely to contain most of
the source counts.  An example event is shown in Figure \ref{fig:triggerNprob} (Top Panel).

\noindent 1. For every possible source extraction window t1--t2, by examining the cumulative distribution
of detected counts in a light curve with 10ms bins, we record each interval of duration $\Delta t$ [s] with signal-to-noise 
ratio $S/N$ over threshold $S/N_{\rm min} = {\rm MIN} ( \sqrt{\Delta t} , 5 )$.
This trigger threshold suppresses the detection of entire emission episodes lasting
longer than 25s.  This is to avoid contamination due to count rate fluctuations 
that sometimes occur at the start or end of data acquisition due to the spacecraft
slew.  Low $S/N$ and long emission episodes
are still detected, provided they are comprised of shorter regions, because:

\noindent 2. We sort the triggers and dump temporal overlaps with lower $S/N$.  The burst
region is defined as the interval containing all surviving intervals.  For the example
shown in Figure \ref{fig:triggerNprob} (Top Panel), the algorithm recovers four temporally
separate triggers over threshold.

\noindent 3. We allow the endpoints of this region to extend slightly outward to allow for the presence of a low $S/N$ rise
or tail.  With binsize $0.01~dt_{S/N}$, where $dt_{S/N}$ is the duration of the time window containing the maximal
$S/N$ detection, we form a binned light curve and denoise the binned light curve with Haar wavelets \citep[e.g.,][]{kol00}.
The start (or end) of the initial burst region is allowed to extend by one additional bin for a total extension of
$n_{\rm extend}$ bins as long as the $S/N$ of the denoised lightcurve in that bin is $>0.1~\sqrt{n_{\rm extend}}$, where 0.1 is 
the typical root-mean-square background noise fluctuation after denoising.

This final region is fixed and used for the timing and spectral analyses discussed below.
In three cases (GRBs 060218, 061027, and 070126), the above procedure fails to detect a trigger
and we must decrease the threshold in step 1 to $S/N_{\rm min} = 3$.

\subsection{Burst Duration Estimates}
\label{sec:durs}

Using the burst intervals defined above for each event, we form the cumulative distribution of source
counts and record the time values according to when a fraction 5, 25, 75, and 95\% of the total
counts arrive relative to the start of the burst interval.  The difference between the 75 and
25 percentile time defines the burst $T_{50}$ duration, while the difference between the 95 and
5 percentile time defines the burst $T_{90}$ duration.  We also determine a measure of duration
$T_{r45}$ according to the prescription of \citet{reichart01}.  We also report the ratio of the 
peak rate Rate$_p$ (in a time bin of width $0.01~dt_{S/N}$) over the total source counts (Cts).
This is used to below to approximately relate the burst fluences reported in Table 2 to peak fluxes.
We determine errors on each measured
duration by performing a bootstrap Monte Carlo \citep[e.g.,][]{lupton93}, using the observed Poisson
errors on the observed count rate.  These duration values as well as the time region
and $S/N$ ratio of the highest $S/N$ trigger for each burst are given in Table 1.

$T_{90}$ durations are strongly dependent on the choice of burst start and stop
times, which are typically set by hand \citep[e.g.,][]{paciesas99}.  We note the following
loose consistency with the $T_{90}$ values reported by the {\it Swift}~team without
uncertainties on their
public webage\footnote{http://swift.gsfc.nasa.gov/docs/swift/swiftsc.html}.
Less than half (39\%) of our $T_{90}$ values are consistent (1-sigma level) with those
reported on the {\it Swift}~webpage, assuming a 10\% error for the {\it Swift}~Team
values.  At the 3-sigma level, the consistency is 67\%.  Although individual
values are inconsistent, it is important to note that we cannot reject the
hypothesis that our $T_{90}$ distribution \citep[see, also,][]{jason_paper} is 
consistent with that of the {\it Swift}~Team (Kolmogorov-Smirnov (KS) test 
$P_{\rm KS} = 0.7$).
                                                                                                              
\subsection{Caveats, Manual Burst Region Edits}
\label{sec:caveats}

We separate short and long durations GRBs \citep[e.g.,][]{kouv93}
here using a cutoff duration $T_{90}<3$s.  The
details of how our durations change with energy band and redshift are discussed in
\citet{jason_paper}.
In some cases, our automated algorithm detects a faint and long tail following some short GRBs.
For GRB~050724, for example, there is a broad ($dt=20.7$ s) and late bump at 74 s after the  main pulse ($dt=0.54$ s, $S/N=20.8$).
For GRB~061006, the automated region selection above finds a broad $\approx 60$ s \citep[see, also,][]{krimm06} burst region
due to a broad pulse with $S/N=14.7$.  This lies under and after a narrow pulse of
$S/N=42.6$ and $dt=0.82$ s.  The ratio of the durations of these pulses
($\approx 40$ and 70, respectively) are $\gtrsim 3\sigma$ outliers with respect to the ratios found for all other BAT events.  For this reason, we
present in Table 2 spectral fits for both the full burst region and for the narrow pulse.  We do this also for
GRB~051227.  We also conservatively label these events as ``short-duration'' events for exclusion in
the analyses in Section \ref{sec:discuss}.

The portion of of GRB~060124 which we analyze is only the pre-cursor to a much longer event \citep[see, e.g.,][]{romano06b}.
The precursor is a factor $\approx 15$ fainter than the large flare which
occurs $\approx 500$s later and for which only a light curve data are available.

Similarly we do not analyze the flux contribution to the unusual GRB~060218 \citep[e.g.,][]{campana06a} after $t\approx 300$s.

\subsection{Experimental Determination of the Detector Threshold}
\label{sec:thresh}

Plotting various observed quantities derived from the BAT spectra and light
curves, we notice a strong correlation between the photon fluence and duration (e.g.,
Figure \ref{fig:thresh}, Top Right).  This correlation has also been noted by \citet{berger07s} for energy
fluence in the case of short-duration GRBs only.  \citet{lbp00} discuss a similar correlation
found for pulses in {\it BATSE}~bursts, which is unlikely to be due to cosmological effects.

The fluence and duration in the {\it Swift}~sample are best fit
by a powerlaw with index consistent with one half, which is suggestive of a
detector threshold at the Poisson level.  It is reasonable that the BAT
detector could perform at or near the Poisson level over a wide range of
burst durations due to the detector's capacity to trigger on images (demasked
light curves).  A precise determination of the threshold, which is beyond
the scope of this work, would involve modelling
the satellite triggering algorithm and observational efficiency and also accounting for the sensitivity by the
detector at different field angles for incident photons distributed in energy according
to the true burst spectrum.
We are interested here in obtaining an approximation to this
threshold in terms of our best-fit values for detector independent quantities.

The fluence--duration correlation is likely due in part to the both shape of the typical
GRB spectrum and also due to an intrinsic decrease in the number of bright relative
to faint events.  To test whether a truncation of the lowest fluence values by
the detector threshold also contributes to the correlation, we plot the histogram of
photon fluence over the square root of the $T_{90}$ duration (Figure \ref{fig:thresh}, Top Left).
There is a narrow clustering of values, and we find that $>90$\% of events
have $n_{\rm bol}/\sqrt{T_{90}}> 3$ ph cm$^{-2}$ s$^{-0.5}$.  Also,
characteristic of a threshold, the observed $S/N$ of the maximal $S/N$ image trigger correlates
tightly and linearly with $n_{\rm bol}/\sqrt{T_{90}}$ (Figure \ref{fig:thresh}, Bottom Left).
This clustering does not tighten if we consider a threshold in terms of peak photon rate instead of
fluence over root time, as is typically the case for GRBs which fade rapidly in time \citep[e.g.,][]{band03}.

We find that the threshold in $n_{\rm bol}/\sqrt{T_{90}}$ corresponds to
an $\approx 5\sigma$ detection threshold.  The scatter around this best fit log-log line is
$\sigma=0.52\pm0.05$, determined using equation (\ref{eq:scat}).  Hence, the threshold
estimator traces the actual threshold (as proxied by the observed $S/N$) to $\approx 50$\% accuracy.
There is no significant decrease in the scatter
in Figure \ref{fig:thresh} (Bottom Left) or significant increase in the tightness of the histogram
in Figure \ref{fig:thresh} (Top Left) if we attempt to include $E_{\rm pk,obs}$ (to some power) in
the threshold estimate.  We note that our value of $(n_{\rm bol}/\sqrt{T_{90}})_{\rm thresh}$
is closely consistent with the value estimated prior to {\it Swift}~of $\approx 1$
ph cm$^{-2}$ s$^{-1}$ peak rate (1-$10^3$ keV) by \citet{band03}, after accounting for a slight increase
due to a typical $\sqrt{T_{90}}\approx 3$ s$^{0.5}$.  The \citet{band03} threshold is
also nearly independent of $E_{\rm pk,obs}$.

\section{Spectral Fitting}
\label{sec:spec}

We employ in parallel two spectral modelling approaches.  The first is
a classical frequentist approach that will be familiar to experienced
users of the software package XSPEC \citep{arnaud96}.  As we
describe in the next subsection, we fit the data with the simplest 
of three possible models.  We then derive confidence intervals by considering
random realizations of the data given the best-fit model 
for each model parameter constrained by the best-fit model.  This approach
turns out to be very limited for {\it Swift}~events, due to the narrow energy
bandpass of the BAT instrument.  In particular it is possible to measure a $\nu F_{\nu}$
spectral peak energy $E_{\rm pk,obs}$ for only about one third of the events in the sample.
A more powerful Bayesian approach assumes that the each burst spectrum has
an intrinsic spectrum containing the interesting $E_{\rm pk,obs}$ parameter, and we
derive the probability distribution for that parameter given the 
data.  We show below that prior information can be exploited to derive 
limits on $E_{\rm pk,obs}$ (and the burst fluence) even for cases where
$E_{\rm pk,obs}$ is well above the detection passband.

\subsection{Model Fitting 1: Frequentist Approach}
\label{sec:freq}

We fit the time-integrated BAT spectra in the 15-150 keV band by forward folding 
an incident photon spectrum through the detector response.  The resulting counts model
is called $m(\vec \theta)$ and is a function of the parameters $\vec \theta$.  We find
the best-fit model by minimizing:
\begin{equation}
  \chi^2 = \sum_i (y_i-m_i(\vec \theta))^2/\sigma_i^2,
\end{equation}
where $y_i$ is the count rate per energy in energy $E$ bin $i$ and $\sigma_i$ is the uncertainty
(estimated from the source and background data) in $y_i$.   To avoid falling into local
$\chi^2$ minima, all minimization is done using a
downhill simplex algorithm \citep[e.g.,][]{press92} instead of the default XSPEC Marquardt algorithm \citep{arnaud96}. 

We consider three possible models of increasing complexity to fit the time-integrated BAT
spectra.  These are a simple powerlaw, a powerlaw times an exponential cutoff, and 
a smoothly-connected broken powerlaw.  The final model is the GRB Model (GRBM) of \citet{band93}.
We force this model to have a peak in the $\nu F_{\nu}$ spectrum in $E\in (0,\infty )$ 
by requiring that
the low energy photon index $\alpha>-2$ and the high-energy photon index $\beta<-2$.
Identifying the exponential times powerlaw model with the low energy portion of the GRBM
spectrum, we require that the photon index for this model also satisfy $\alpha>-2$.

As we step from one model to the next, we add one additional parameter to the fit.  Because
the models are nested, the improvement in $\chi^2$ with each new parameter is distributed
approximately as $\chi^2_{\nu}$ with $\nu=1$ degrees of freedom \citep[e.g.,][]{protassov02}.  We only
allow the step to a more complex model if the change in $\chi^2$ corresponds to a $>$90\%
confidence improvement in the fit.  
Errors on the parameters $\vec \theta$ are reported in Table 2 and are found from $\chi^2_{\rm min}$ in 
the vicinity of the global minimum as described in, e.g., \citet{cash76}.

The middle panel of Figure \ref{fig:triggerNprob}
shows an example spectrum which is well fit by a powerlaw.  All BAT bursts are adequately 
fit by one of the three models (Table 2).

In the 63\% of cases where the data are adequately fit by a simple powerlaw model only, we also
calculate a limit on $E^{\rm freq.}_{\rm pk,obs}$ as follows.  If the photon index is more negative than
$-2$ at 90\% confidence, we use the constrained Band formalism \citep{taka04} to derive an $E^{\rm freq.}_{\rm pk,obs}$ upper limit.
If the photon index is greater than $-2$ at 90\% confidence, we derive an $E^{\rm freq.}_{\rm pk,obs}$ lower limit by fitting
an exponential times powerlaw model.  We warn the reader that lower and upper limits, respectively, 
on $E^{\rm freq.}_{\rm pk,obs}$ are undefined in these cases.  Also, the probabilities associated with all
quantities become poorly defined if the best-fit models have $\beta\approx -2$ or $\alpha \approx -2$,
due the discontinuity in $\chi^2$ at these values \citep[see, e.g.,][]{protassov02}.

\subsection{Model Fitting 2: Bayesian Approach}
\label{sec:bayesfit}

In the discussion below, we will be primarily interested in determining burst energetics
via $E_{\rm pk,obs}$ and the bolometric GRB energy fluence.  Because these quantities are poorly
defined (if at all) for many BAT events in the frequentist approach, we consider also a more powerful Bayesian
approach.  The likelihood of the model given the data is 
\begin{equation}
  P(\vec Y|\vec \theta) \propto \exp{(-\chi^2/2)}
\end{equation}
From Bayes rule \citep[e.g.,][]{greg05} the posterior distribution gives the probability of the model, given
the data
\begin{equation}
  P(\vec \theta |\vec Y) \propto P(\vec Y|\vec \theta) P(\vec \theta),
\end{equation}
where $P(\vec \theta)$ in the prior probability on the model.  We assume below that the
prior can be broken into four multiplicative terms, one for each of the GRBM parameters.  The posterior probability distribution for
a given parameter $\theta_i$ is found by marginalizing $P(\vec \theta |\vec Y)$ over
the other parameters.

The power in the Bayesian approach comes from its capacity to allow us to incorporate pre-{\it Swift}~knowledge
of GRB spectra into our model fitting through the prior.  Observations of thousands of GRBs by
{\it BATSE}~\citep[e.g.,][]{preece00} strongly limit the range of likely GRBM parameters.  

Most importantly (for {\it Swift}),
the {\it BATSE}~distribution in $E_{\rm pk,obs}$ falls off sharply above 300 keV \citep{preece00,kaneko06}.  This leads us to
the following prior on $ln[E_{\rm pk,obs}]$, ignoring the normalization:
\begin{multline}
P( ln[E_{\rm pk,obs}] ) \propto \exp{\left \{ -0.5\left[ln \left( { E_{\rm pk,obs} \over 300 ~{\rm keV}} \right) \right]^2/\sigma^2_{lE_p} \right \}},
\end{multline}
for $E_{\rm pk,obs}>300$ keV, with $\sigma_{lE_p}=4/5$.  We assume a uniform distribution in $ln[E_{\rm pk,obs}]$ below 300 keV
instead of the cutoff observed by {\it BATSE}~due to the high-energy bandpass of that instrument
and to the discovery of X-ray Flashes \citep[XRFs;][]{heise2000} which extends the distribution to low $E_{\rm pk,obs}$.
We assign zero probability to $E_{\rm pk,obs}<1$ keV and $E_{\rm pk,obs}>10^4$ keV.

\citet{kippen02} \citep[also,][]{barraud03} show that the photon indices for XRFs are consistent with 
those found for GRBs.
We assume the low energy powerlaw index distribution from {\it BATSE}~\citep[e.g.,][]{kaneko06}:
\begin{equation}
P( \alpha) \propto \exp{(-0.5(\alpha-\alpha_{\rm pk})^2/\sigma^2_{\alpha})},
\label{eq:alpha}
\end{equation}
with $\alpha_{\rm pk}=-1.1$ and $\sigma_{\alpha}=0.25$.

There is evidence that the high-energy index distribution broadens with the inclusion of XRFs \citep[see,][]{taka05}.
To be conservative, we assume only the peak of the {\it BATSE}~distribution $\beta_{\rm pk}=-2.3$.  We use
the maximum entropy \citep[e.g.,][]{greg05} prior for a distribution of known mean:
\begin{equation}
P( \beta ) \propto \exp{(-\beta/\beta_{\rm pk})}.
\end{equation}

Finally, we assign equal probability per logarithmic interval to the model normalization, taken to be the fluence
in the $1-10^4$ keV band (host frame, or source frame if redshift unknown) \citep[see, also,][]{amati02}.  We 
truncate this prior below $10^{-10}$ erg cm$^{-2}$ so that the
integral over the model normalization remains finite.  The specific value of this truncation is unimportant.  We find 
identical results if we truncate instead at $10^{-50}$ erg cm$^{-2}$.

\subsection{Most Probable Values, Samples, and Confidence Intervals}

We find that the $E_{\rm pk,obs}$ marginal posterior probability distributions are typically broad and asymmetrical.
We calculate these distributions explicitly for each event by integrating $P(\vec \theta |\vec Y)$ analytically
over model normalization and numerically over $\alpha$ and $\beta$.  The 2-dimensional numerical integration
is done via 10-point Gaussian quadrature \citep{press92}.  At each step, we fit for and concentrate the integration in the
region of maximal $P(\vec \theta |\vec Y)$.  

An example posterior probability curve is plotted in
the bottom panel of Figure \ref{fig:triggerNprob}.  With the adopted prior above, we recover an $E_{\rm pk,obs}$
value which is well above the {\it Swift}~BAT bandpass and also consistent with Konus-Wind measurements \citep{gol06} in the 20--$2 \times 10^4$ keV band.

Figure \ref{fig:triggerNprob} also shows how the $E_{\rm pk,obs}$ determination changes as we relax our priors.  Quadrupling the 
dispersion $\sigma_{\alpha}$ in the $\alpha$ prior has little effect at these high energies.  However, if we discard
the prior on $ln[E_{\rm pk,obs}]$ above 300 keV, then we are only able to derive a lower limit on $E_{\rm pk,obs}$ as in the frequentist
approach ($E_{\rm pk,obs}>390$ keV; Table 2).  At low energies, and analogous to the constrained Band formalism \citep{taka04},
the prior on $\alpha$ helps to break the degeneracy 
between fitting a powerlaw spectrum associated with either the low-energy or the high-energy portion of the GRBM.

To describe the joint posterior probability distribution in $E_{\rm pk,obs}$ and fluence $E_{\rm iso}$, we obtain
$10^3$ samples via a Markov Chain Monte Carlo (MCMC) simulation.  We first draw $10^3$ samples from
the marginal $E_{\rm pk,obs}$ posterior distribution (tabulated as discussed above).  For each $E_{\rm pk,obs}$, we determine the mode
of P($E_{\rm iso}$,$E_{\rm pk,obs}|\vec Y$) and the curvature at
the mode.  These define a Gaussian sampling distribution from which we take
five Metropolis-Hastings steps \citep[e.g.,][]{greg05} in a random walk.
The last $E_{\rm iso}$ value
is saved, and the process is repeated $10^3$ times to store $10^3$ $E_{\rm pk,obs}$--$E_{\rm iso}$ pairs.
Samples from the posterior distributions for each GRB with measured redshift are plotted in Figure \ref{fig:amati}.
We use the
$E_{\rm pk,obs}$ and $E_{\rm iso}$ samples to report the most probable values and intervals containing
90\% of the posterior probability for all 218 bursts in Table 2.

\subsection{$E_{\rm pk,obs}$ Constraints for Powerlaw Events}
\label{sec:powerlaws}

The referee has noted that our most-probable $E_{\rm pk,obs}$ values for events adequately
fit by simple powerlaws correlate tightly with the best fit powerlaw indices $\alpha_{\rm best-fit}$ 
\citep[Figure \ref{fig:alpha_ep}; also,][]{taka06}.  Indeed, because we wish to measure $E_{\rm pk,obs}$ from data
which only constrain $\alpha_{\rm best-fit}$, this had better be the case.  How
does the correlation arise?

Half of the events, or about 30\% of the total sample, produce the
tightest region of the correlation and have $\alpha_{\rm best-fit}<-1.6$.  Given 
our prior on the GRBM $\alpha$, these $\alpha_{\rm best-fit}$
are $2\sigma$ unlikely to be associated with the low energy index.  Instead, the most-probable
model steepens the index by placing $E_{\rm pk,obs}$ just above, in, and then below the BAT bandpass 
\citep[see, also,][]{cabrera07}.
We should not, therefore, interpret $\alpha_{\rm best-fit}$ as the GRBM $\alpha$ in this
regime.  However, we can also be confident that the data are strongly influencing $E_{\rm pk,obs}$.

The remaining half of events have $\alpha_{\rm best-fit}$ which could be associated with the GRBM $\alpha$.
Above 200 keV, the steepening of $\alpha_{\rm best-fit}$ relative to $\alpha$ is comparable to the
breadth of the $E_{\rm pk,obs}$ prior, and this makes the prior on $\alpha$ relatively unimportant in this
regime (e.g., Figure \ref{fig:triggerNprob}, Bottom Panel).
Because the exponential cutoff in the GRBM $E_{\circ}=(2+\alpha)E_{\rm pk,obs}$, the hardest events will
still be most sensitive to low $E_{\rm pk,obs}$ values and will lead to tighter lower limits than for the softer
events.  The prior on $E_{\rm pk,obs}$ truncates the probability at high $E_{\rm pk,obs}$, and we continue
to see a correlation, although with more scatter.  Figure \ref{fig:alpha_ep} shows how the $E_{\rm pk,obs}$ upper 
limits effectively account for possible large $E_{\rm pk,obs}$ values in the {\it BATSE}~sample.

The average $E_{\rm pk,obs}$ for a large number of events in the regime $\alpha_{\rm best-fit}>-1.6$ is expected,
therefore, to be unbiased with respect to {\it BATSE}~GRBs with $E_{\rm pk,obs}> 200$
keV.  Also, the uncertainty in our estimate should account for the population $E_{\rm pk,obs}$ variations at high
$E_{\rm pk,obs}$.  Therefore, our $E_{\rm pk,obs}$ should still be useful for population studies
(e.g., Sections \ref{sec:amati},\ref{sec:other}).  Our error regions are also likely to contain the true
$E_{\rm pk,obs}$ for a given event.

Whether the most likely $E_{\rm pk,obs}$ for an individual burst closely corresponds to
the
true $E_{\rm pk,obs}$ will depend on whether or not $\alpha$ tends to be shallow for high $E_{\rm pk,obs}$.
There is a weak but signifcant correlation ($\tau_K=0.18$, $5.1\sigma$) in GRBM fits to the \citet{kaneko06} 
{\it BATSE}~sample, which indicates that this may be the case for some events.
Contemporaneous measurement at energies above the BAT bandpass provide
a direct test.

\subsection{Comparison to Konus-Wind and Suzaku Measurements}
\label{sec:kw}

For 75 events in Table 2, we are able to determine lower and upper 90\% confidence intervals for $E_{\rm pk,obs}$
using the classical frequentist approach (Figure \ref{fig:catalog}; Top Left).
The sample mean is $E^{\rm freq.}_{\rm pk,obs} = (79\pm 6)$ keV.  We find consistent and unbiased $E_{\rm pk,obs}$ estimates
from the Bayesian approach ($E_{\rm pk,obs} = (83\pm 6)$ keV).  There is no strong evidence that the distributions
are inconsistent ($P_{\rm KS}=0.88$).  How
does the Bayesian approach fare at higher energies where the spectra are typically acceptably modelled by powerlaws?

Comparing our {\it Swift}~numbers to values from 27 observations
by Konus-Wind reported in the Gamma-ray bursts Coordinates Network (GCN) circulars \citep[e.g.,][]{gol06},
we find no evidence for bias in either our $E_{\rm pk,obs}$ or our
$S_{\rm bol}$ determination (Figure \ref{fig:sks}).
The sample means for both quantities are consistent at the $<1\sigma$ level
($\log_{10}{(S_{\rm bol,KW})} = -4.62 \pm 0.08$ versus $\log_{10}{(S_{\rm bol,Sw})} = -4.62 \pm 0.09$
and
$\log_{10}{(E_{\rm pk,obs,KW})} = 2.52 \pm 0.06$ versus $\log_{10}{(E_{\rm pk,obs,Sw})} = 2.47 \pm 0.06$).
There is no evidence from a KS-test ($P_{\rm KS}=$1.0, 0.9, for the fluence and $E_{\rm pk,obs}$ comparisons,
respectively) that the distributions are different.
Additionally, we note that there are very few discrepancies.  We find 
that $>85$\% of either our $E_{\rm pk,obs}$ values or our $S_{\rm bol}$ values 
are consistent within our estimated 90\% confidence errors.  

This agreement is remarkable considering that the Konus-Wind spectral fits are only the preliminary 
fits reported to the GCN.  For seven of the events, Suzaku measurements are also reported in the GCN
\citep[e.g.,][]{hong07}.  We find no evidence for bias when comparing our {\it Swift}~values instead to these (Figure \ref{fig:sks}).

To check that we are measuring $E_{\rm pk,obs}$ values above the BAT bandpass rather than simply assigning these
all the same $E_{\rm pk,obs}$, we fit a powerlaw to the {\it Swift}~and Konus-Wind data in Figure \ref{fig:sks} (Left).
To be conservative, we exclude the 8 points below 200 keV.  For most (15 of 19)
of these events, the {\it Swift}~data are adequately fit by a powerlaw, and
only one event has $\alpha_{\rm best-fit}<-1.6$.  The
remaining 4 of 19 events have weakly constrained $E^{\rm freq.}_{\rm pk,obs}$ 
values.  The joint $E_{\rm pk,obs}$ data are
fit by a powerlaw with index ($0.6\pm0.2$) greater than zero at 
$\approx 5\sigma$ confidence.  

Although this indicates information content
in our $E_{\rm pk,obs}$ values, an index less than unity indicates a tendency to underestimate 
$E_{\rm pk,obs}$ at high $E_{\rm pk,obs}$.  Our prior appears to lead to underestimates of large $E_{\rm pk,obs}\gtrsim 2$ MeV by
a factor $\gtrsim 2$.  We note that our estimates remain consistent within 
errors, that $E_{\rm pk,obs}$ understimates will be conservative as regards the analyses below, and that such very high $E_{\rm pk,obs}$ 
events are rare in the {\it BATSE}~sample.

\subsection{Comparison to BATSE Measurements}

Figure \ref{fig:catalog} displays our fluence and $E_{\rm pk,obs}$ estimates.  In the lower-right panel,
an excess of low-$E_{\rm pk,obs}$ events is present relative to the $E_{\rm pk,obs}$ distributions determined
by \citet{kaneko06} for bright {\it BATSE}~GRBs.  A similar effect is present in the {\it HETE-2}~catalog
\citep{taka05}.  We also plot the best-fit Gaussian to the $E_{\rm pk,obs}$ data including errors (Section \ref{sec:amati}).
There is marginal evidence for a shift in the peak of the $E_{\rm pk,obs}$ distribution. The
prior we assume on $E_{\rm pk,obs}$ (from {\it BATSE}) is at least partially responsible for this effect.
Further analyses and comparisons of our $E_{\rm iso}$ and $E_{\rm pk}$ estimates to those found for previous
GRBs, also using our XRT analyses \citep{butler07a,butler07b},  are presented in \citet{dans_paper}.

\section{Discussion}
\label{sec:discuss}

\subsection{$E_{\rm iso}$--$E_{\rm pk}$ Correlation and Intrinsic Scatter}
\label{sec:amati}

In this section, we use the $10^3$ pairs of $E_{\rm iso}$--$E_{\rm pk}$ samples accumulated above, which
fully account for correlations between the GRBM parameters for each of 77 bursts with measured redshift,
in order to test the well known burst-to-burst relation $E_{\rm pk} = (1+z)E_{\rm pk,obs} \approx K (E_{\rm iso}/[10^{52}{\rm ~erg}])^{\eta}$ 
keV \citep[e.g.,][]{lpm00,amati02}.
The samples, aside from the most peripheral 10\%, are plotted in Figure \ref{fig:amati} along with the best fit
powerlaw we derive below.  These are publicly available from our webpage\footnote{http://astro.berkeley.edu/$\sim$nat/swift}
and should be used in place of the best-fit values (Table 2) when fitting models to the data.  As we illustrate below, these samples
can be used to rigorously determine the normalization, slope, and scatter, even when large
measurement errors are present.

First it is interesting to know what fraction of the total number of events have at least $10^2$ samples on the opposite
side of the best fit line of \citet[][i.e. $K=90$, $\eta=0.49$]{amati02} from the main mass of samples.  This provides a measure of consistency at the 90\% confidence
level.  To be conservative, we allow the \citet{amati02} relation to have a logarithmic dispersion of $\pm0.7$ (i.e., 0.3 dex).  
We also toss out strong outliers from our sample.
We ignore the short-duration and underluminous events marked in yellow and cyan in the Figure \ref{fig:amati}.  We use an ad hoc (and admittedly
circular) definition of underluminous: the $E_{\rm iso}$ is 100 or more times fainter than that expected from the \citet{amati02}
relation, given the best-fit $E_{\rm pk}$.  The underluminous events with long durations are GRBs 051109B, 060218, and 060614.
After removing the underluminous and short duration events and retaining only those
marked with red circles in Figure \ref{fig:amati},
we find that 41\% ($26/63$) of {\it Swift}~events are inconsistent with the \citet{amati02} powerlaw at the 
90\% confidence level.  

We now determine the best $E_{\rm iso}$--$E_{\rm pk}$ powerlaw relation for the {\it Swift}~data.  
For each event $j$, the $10^3$ samples represent the posterior probability $P(E_{j,{\rm iso}},E_{j{\rm pk}}|\vec Y)$ (Section \ref{sec:bayesfit}).
Using the $(E_{\rm iso},E_{\rm pk})$ values corresponding
to the posterior peak in each event (Table 2), there is evidence for a strong correlation (Section \ref{sec:amati_cor}).
What are the values of $K$ and $\eta$ describing this correlation via a powerlaw fit, and what is the true scatter around this fit?

For simplicity in notation we write $x_j = ln(E_{j,{\rm iso}}/[10^{52} {\rm ~erg}])$ and $y_j=ln(E_{j,{\rm pk}})$.
We now assume for each burst a prior between $x_j$ and $y_j$ representing a powerlaw relation between $E_{\rm iso}$ and $E_{\rm pk}$:
\begin{multline}
  P(x_j,y_j|k,\eta,\sigma_A) \propto \\ \exp{( -0.5[y_j-k-\eta x_j]^2/\sigma^2_A )}/\sigma_A,
  \label{eq:allpost}
\end{multline}
where $\sigma_A$ allows for an {\it intrinsic}~scatter in the correlation.  By inverting the data in this fashion to determine
the intrinsic scatter (rather than just assuming that the observed scatter is the intrinsic scatter), we rigorously
account for the large $E_{\rm pk,obs}$ uncertainties arising from the narrow BAT bandpass.
The parameter $\sigma_A$ also plays an important role in allowing the powerlaw model to acceptably fit the data,
as we discuss in Section \ref{sec:cabrera}.

Equation (\ref{eq:allpost}) multiplies the posterior $P(x_j,y_j|\vec Y)$
for each event to form the posterior probability of $k$, $\eta$, and $\sigma_A$ for that event.  The posterior considering all $N$
events is then:
\begin{multline}
  P(k,\eta,\sigma_A,x_j,y_j|\vec Y) \propto ~ \sigma^{-N-1}_A \prod_{{\rm event}~j} P(x_j,y_j|\vec Y) \\
 \times \exp{( -0.5[y_j-k-\eta x_j]^2/\sigma^2_A )}.
 \label{eq:p1}
\end{multline}
Here we have included a $1/\sigma_A$ prior on $\sigma_A$ (i.e., equal probability per logarithmic interval or scale invariance).  We assume
uniform priors on $k$ and $\eta$.  Because $P(x_j,y_j|k,\eta,\sigma_A)=P(y_j|x_j,k,\eta,\sigma_A)P(x_j|k,\eta,\sigma_A)$, we are effectively setting
a uniform prior on $x_j$ as well.

We wish now to marginalize over the $(x_j, y_j)$.  This can be accomplished by Monte Carlo integration using the
$10^3$ samples $i$ for each event:
\begin{multline}
  P(k,\eta,\sigma_A|\vec Y) \propto ~ \sigma^{-N-1}_A \\
  \times \prod_{{\rm event}~j} \left \{ \sum_{{\rm samp}~i~} \exp{( -0.5[y_{i,j}-k-\eta x_{i,j}]^2/\sigma^2_A )} \right \},
\end{multline}
where $x_{i,j}$ designates the $i$th sample of $x_j$.
Because the $P(x_j,y_j|\vec Y)$ are independent, we can carry out
the product before the sum in Equation \ref{eq:p1}, provided we randomize away any sorting that may have occurred in the tabulation 
of the $(x_{i,j},y_{i,j})$. 
Defining $\Theta_i \equiv \sum_j [y_{i,j}-k-\eta x_{i,j}]^2$, we have:
\begin{multline}
P(k,\eta,\sigma_A|\vec Y) \propto ~ \sigma^{-N-1}_A \left \{ \sum_{{\rm samp}~i~} \exp{( -0.5\Theta_i/\sigma^2_A )} \right \}.
\end{multline}
If we define the following statistics for the set of $i$th samples averaged over events $j$, 
$mx_i={<x_{i,j}>_j}$, $my_i={<y_{i,j}>_j}$
$vx_i={<x^2_{i,j}>_j - mx^2_j}$,
$vy_i={<y^2_{i,j}>_j - my^2_j}$,
$cov_i={<x_{i,j}y_{i,j}>_j} - mx_jmy_j$, we can marginalize to find:
\begin{multline}
P(k,\eta|\vec Y) \propto \sum_{{\rm samp}~i} [k^2 + 2 k \eta ~mx_i - 2 k ~my_i + \\
  \eta^2 (vx_i+mx^2_i) - 2 \eta (cov_i+my_i ~mx_i) + (vy_i+my^2_i)]^{-N/2},
 \label{eq:ka}
\end{multline}
\begin{multline}
   P(k|\vec Y) \propto \sum_{{\rm samp}~i} (vx_i+mx_i^2)^{(N-2)/2} \\ [(vy_i+my_i^2) (vx_i+mx_i^2)- (cov_i+ 
   mx_i my_i)^2 - \\ 2 k (my_i ~vx - mx_i ~cov_i) + k^2 ~vx_i )]^{-(N-1)/2},
  \label{eq:k}
\end{multline}
\begin{multline}
   P(\eta|\vec Y) \propto \sum_{{\rm samp}~i} ( \eta^2 ~vx_i - 2 \eta~cov_i+vy_i )^{-(N-1)/2}, ~{\rm and}
  \label{eq:b}
\end{multline}
\begin{multline}
   P(\sigma_A|\vec Y) \propto \sigma^{-N+1}_A \\ \times \sum_{{\rm samp}~i} vx_i^{-1/2} \exp{(-0.5 N [vy_i-cov_i^2/vx_i]/\sigma^2_A)}.
   \label{eq:scat}
\end{multline}
In a similar fashion to the derivation of these formulae, it is interesting to know the intrinsic distribution
of individual parameters if we assume that this distribution is a Gaussian, i.e.,
\begin{multline}
  P(x_j|\vec Y) \propto \exp{( -0.5[x_j-x_{\circ}]^2/\sigma^2_{\rm x} )}/\sigma_{\rm x}.
  \label{eq:gauss}
\end{multline}
As in Equation (\ref{eq:ka}), 
\begin{multline}
P(x_{\circ}|\vec Y) \propto \sum_{{\rm samp}~i} [(x_{\circ}-mx_i)^2 + vx_i]^{-N/2}.
  \label{eq:onex}
\end{multline}
As in Equation (\ref{eq:scat}), 
\begin{multline}
   P(\sigma_{\rm x}|\vec Y) \propto ~ \sigma^{-N}_{\rm x} \sum_{{\rm samp}~i} \exp{(-0.5 N vx_i/\sigma^2_{\rm x})}.
   \label{eq:scatx}
\end{multline}

Figure \ref{fig:amati_prob} shows Equations (\ref{eq:ka}, \ref{eq:k}, and \ref{eq:b}) for the 63 events marked by red circles in Figure \ref{fig:amati}.
We find that the posterior (Equation \ref{eq:ka}) peaks at $\log_{10}{(K)}=2.35\pm 0.09$
and $\eta = 0.47\pm0.08$.  
This is inconsistent at the $>5\sigma$ level with all pre-{\it Swift}~curves.  There is also large {\it intrinsic}~scatter in the
relation $\sigma_A = 0.7\pm0.1$ ($0.30\pm0.04$ dex), given the {\it Swift}~data.  

Comparatively, the observed scatter (Equation \ref{eq:scat} for just the 
best fit $E_{\rm pk}$ and $E_{\rm iso}$ values; Table 2) is 0.46 dex.
This latter value is an unfair estimate of the true scatter, because it is contaminated by the relatively weak $E_{\rm pk,obs}$ determinations (Figure \ref{fig:amati})
due to the narrow BAT bandpass.  The intrinsic scatter we calculate 
is far larger than the pre-{\it Swift}~estimate of $0.14^{+0.3}_{-0.2}$ dex
by \citet{amati06}.
Because $\sigma_A$ refers to the scatter in the logarithm, our value corresponds to a factor of 
2 intrinsic scatter in the powerlaw relation $E_{\rm pk}=K E_{\rm iso}^{\eta}$.

\citet{li2007} observes a possibly significant variation in $K$ and $\eta$
with redshift for the pre-{\it Swift}~sample.  
For the {\it Swift}~sample, we observe no significant evidence for such variations
(Figure \ref{fig:amati_prob}). Also, we note that possible
variations appear to be non-monotonic.
The intrinsic scatter does not vary significantly:
$\sigma_A=0.58^{+0.20}_{-0.14}$ ($z<1.5$),
$\sigma_A=0.53^{+0.24}_{-0.14}$ ($1.5<z<3.0$), and $\sigma_A=0.65^{+0.20}_{-0.16}$ ($z>3$).

There is little evidence for an $E_{\rm pk}$-$E_{\rm iso}$ relation in the {\it Swift}~short-duration GRB
sample, because the relation has large errors 
$\log_{10}{(K)}=2.7\pm 1.0$, $\eta = 0.1^{+0.4}_{-0.5}$
and a large intrinsic scatter $\sigma_A = 1.3^{+0.8}_{-0.5}$.  Also, $\eta$ is consistent with zero.

\subsection{Comparison with \citet{cabrera07} \\ $E_{\rm pk}$-$E_{\rm iso}$ Relation}
\label{sec:cabrera}

As described above, \citet{cabrera07} measure $E_{\rm pk}$ and $E_{\rm iso}$ for 28 {\it Swift}~GRBs
with measured redshift --- a subsample of the full 77 GRB sample considered here.  They account
for the detector-dependent correlation between these quantities for each GRB with a Gaussian approximation.
We have retrieved their best fit $E_{\rm pk}$ and $E_{\rm iso}$ and confidence
regions (their Table 3), drawn $10^3$ samples from the appropriate bivariate-Gaussian distributions for each event, and 
fit the data using Equations (\ref{eq:k}-\ref{eq:scat}).

We find $\log_{10}{(K)}=2.33\pm 0.09$ and $\eta=0.35\pm0.09$ (Figure \ref{fig:amati_prob}), and an
intrinsic scatter $\sigma_A=0.4\pm0.1$ ($0.18\pm0.04$ dex).  
These parameters are
closely consistent with those that we derive above for our full sample, although $\sigma_A$ is larger for the full sample.
We find identical maximum posterior values and confidence regions 
with the {\tt linmix\_err} MCMC regression tool \citep{kelly07} in IDL\footnote{http://idlastro.gsfc.nasa.gov}, which sets different priors for the $E_{{\rm iso},j}$ 
and $\sigma_A$.  

Because \citet{cabrera07} employ a Gaussian error approximation, a powerlaw fit to their $E_{\rm pk}$ and $E_{\rm iso}$ values
can also be obtained from a simple $\chi^2$ minimization procedure.  Following \citet{press92}, we minimize
\begin{multline}
\chi_p^2 = \sum_i { (y_i-k-\eta x_i)^2 \over (\sigma^2_{y_i}+\eta^2\sigma^2_{x_i}-2\rho_i\eta\sigma_{x_i}\sigma_{y_i}+\sigma_A^2) },
 \label{eq:chip}
\end{multline}
where $\sigma_{x_i}$ and $\sigma_{y_i}$ are the errors on $x_i$ and $y_i$, respectively, and $\rho_i$ is the Pearson correlation coefficient
between
$x_i$ and $y_i$ (labeled ``cov'' in Table 3 of \citet{cabrera07}).  Fixing $\sigma_A=0.4$ and minimizing $\chi^2_p$,
we find  $\log_{10}{(K)}=2.32\pm 0.08$ and $\eta=0.32\pm0.19$, for $\chi^2_p/\nu=21.9/26$.  This is consistent with our
Monte Carlo fits above but inconsistent with the Monte Carlo fit reported in \citet{cabrera07}: $\log_{10}{(K)}=2.03\pm 0.01$ and $\eta=0.53\pm0.03$.

For $\sigma_A=0$, the \citet{cabrera07} fit has $\chi^2_p/\nu=445.7/26$.  
It is clear from Figures 4 and 5 in \citet{cabrera07} that their fit does not match the data well and that the
fit errors are under-estimated.
The quoted errors are under-estimated by at least a factor $(\chi^2_p/\nu)^{0.5}\approx 4$.

It is not entirely clear how such a poorly-fitting model was chosen by
those authors.  It is likely that assuming $\sigma_A=0$ precludes finding a statistically acceptable model:
our best-fit model with $\sigma_A=0$ has $\log_{10}{(K)}=2.11$ and $\eta=0.23$ ($\chi^2_p/\nu=366.4/26$).

\subsection{Differences Between the \citet{amati02} Consistent and 
Outlier Samples}
\label{sec:diffs}

Separating the 59\% of events above found to be consistent with the \citet{amati02} relation at 90\% confidence
from those found to be inconsistent, we perform a number of 2-sample KS tests on the observables in Tables 1 and 2.
We adopt a functional definition of inconsistency between these sub-samples: the KS test NULL hypothesis probability
that the sub-samples are drawn from the same parent distribution is $P_{\rm KS}<0.01$.  We find that the sub-sample
distributions goodness of fit $\chi^2/\nu$, $T_{90}$ duration, and fluence are all consistent.  The
redshift and $E_{\rm iso}$ distributions are also consistent.   Contrarily,
the photon fluences are lower on
average by a factor $\approx 2.3$ ($P_{\rm KS}=9.8\times 10^{-3}$) in the observer frame and $\approx 2.7$ in
the source frame ($P_{\rm KS}=1.4\times 10^{-3}$).  Because the energy fluences were consistent, we expect and observe
that the \citet{amati02} inconsistent events
are on average a factor $\approx 2.5$ times harder ($P_{\rm KS}=4.1\times 10^{-4}$) in terms of $E_{\rm pk,obs}$, or 
a factor $\approx 1.5$ in $E_{\rm pk}$ ($P_{\rm KS}=9.3\times 10^{-3}$).  

The differences in $N_{\rm iso}$ and $E_{\rm pk}$ between the two samples are most apparent when we plot the
ratio of these quantities (Figure \ref{fig:amati_thresh}).  Because $E_{\rm iso}\propto E_{\rm pk} N_{\rm iso}$
(e.g., Figure \ref{fig:catalog}, Bottom Left),
a powerlaw relation $E_{\rm peak} \propto E^{0.5}_{\rm iso}$ like that found above for the full {\it Swift}~sample
and by \citet{amati02} for GRBs observed by {\it Beppo-SAX} translates to a line of constant $N_{\rm iso}/E_{\rm pk}$.
We also plot the observed $5\sigma$ detection threshold, determined by scaling the observed $N_{\rm iso}/E_{\rm pk}$
by $5/(S/N)$.  As we discuss in more detail in the next section, a dividing line between events detected by a satellite
of greater sensitivity ({\it Swift}~versus {\it Beppo-SAX}) points to a detector threshold selection effect.

To be quantitative, we perform the following bootstrap simulation
which approximately conserves the local fraction of events per energy flux interval.  In redshift steps $dz=1$, we shuffle
the observed $E_{\rm peak}$, $T_{90}$, and $z$ values and calculate $N_{\rm iso,new} = N_{\rm iso} (E_{\rm pk}/E_{\rm pk,new})/
(T_{90,{\rm new}}/T_{90})$ for each event several times.  We observe that no simulated events above the horizontal line drawn
in Figure \ref{fig:amati_thresh} are lost due to a threshold in $N_{\rm iso}/\sqrt{T_{90}}$ (Section \ref{sec:thresh}).
However, 30\% of events below the line fall below threshold.  If we increase the {\it Swift}~threshold by a factor of 3
\citep[see,][]{band03}
to obtain an approximate {\it Beppo-SAX}~threshold, then 65\% percent of events below the line are lost, while only 2\%
are lost above the line.
If we increase the {\it Swift}~threshold by a factor of 10 --- as is suggested by the \citet{firmani06} relation (Section \ref{sec:lowz})
--- then nearly all (96\%) of events below the line are lost while only 34\% are lost above the line.

An alternative explanation for why our $E_{\rm pk}$--$E_{\rm iso}$ relation is inconsistent with the pre-{\it Swift}~relations
is that the {\it Swift}~GRBs are intrinsically different from pre-{\it Swift}~GRBs.  Most must have $E_{\rm pk}$ values which are 
$(2.5)^2\approx 6$ times harder on average, assuming that they have similar photon fluences.  
We are very confident that our analysis --- which determines $E_{\rm pk,obs}$ for an approximately fixed photon fluence --- could 
not yield an error this large for individual events and certainly not for a large number of events.

Our prior assumptions cannot be the dominant source of inconsistency.
A similar fraction ($\approx 60$\%) of events for both the \citet{amati02} consistent and inconsistent classes
are adequately fit by powerlaws, and the
number of events in each class with $\alpha_{\rm best-fit}<-1.6$ (see Section \ref{sec:powerlaws}) are
comparable.
Moreover, if some events adequately fit by a simple powerlaw in the classical frequentist approach actually did have
extremely high $E_{\rm pk}\gtrsim  1$ MeV, these would be underestimated given our prior, and the actual
values would be {\it more}~inconsistent with the pre-{\it Swift}~relations.

Also, the priors can be discarded, and similar results are found.
Our best-fit $E_{\rm pk}$ and $E_{\rm iso}$ values (Table 2) for the events shared between our and
the \citet{cabrera07} analyses are closely consistent.
The fits lead to a closely consistent $E_{\rm pk}=K E_{\rm iso}^{\eta}$ relation (Section \ref{sec:cabrera}).
If we retain our priors but replace our {\it Swift}~numbers with the 7 closely-consistent $E_{\rm pk,obs}$--$E_{\rm iso}$ pairs available from 
the Konus-Wind sample (Section \ref{sec:kw})
of GRBs with measured redshifts, we also find statistically indistinguishable results.  

The difference in the {\it Swift}~sample cannot be a difference due to the higher redshifts, because the low-$z$ events in the {\it Swift}~sample dominate
the relation and force it to be inconsistent with the pre-{\it Swift}~relation (Figure \ref{fig:amati_prob}).  The
high-$z$ events produce the most consistent $E_{\rm pk}$--$E_{\rm iso}$ relation to that of \citet{amati02}.

\subsection{The $E_{\rm pk}$--$E_{\rm iso}$ Correlation as a Threshold Effect}
\label{sec:amati_cor}

It is apparent from Figure \ref{fig:thresh} (Bottom Right) that the BAT detector threshold leads to 
a strong truncation of detected events with redshift.  We show above how this truncation can help
to narrow the scatter in
a powerlaw $E_{\rm pk}$--$E_{\rm iso}$ relation, making that relation more consistent with a pre-{\it Swift}~relation
found by \citet{amati02}.
Our observed relation corresponds to fainter and harder events, in agreement with indications
from spectral fits to {\it BATSE}~GRBs \citep{np05,bp05} that the relation may in fact be an inequality.

In agreement with previous studies \citep{amati02,amati03,ggl04,fb05,nava06,amati06}, we find that $E_{\rm iso}$ and $E_{\rm pk}$
are tightly correlated (e.g., Figure \ref{fig:amati}).  We find a Kendall's $\tau_K=0.59$ ($6.8\sigma$ significant),
for the best-fit values (Table 2) in 63 bursts.  

A correlation between $E_{\rm pk}$ and $E_{\rm iso}$ could come about in two very
different ways.  First (1), there could be an intrinsic correlation between
these quantities in the source frame, as is widely held \citep[e.g.,][]{cabrera07}.  
Alternatively (2), these quantities could be correlated in the observer frame
(or just narrowly-distributed) and a strong correlation arises when we multiply
both quantities by strong functions of redshift.
The most straight-forward way to rule out (2) in favor of a true source frame
correlation is to
show that there is no strong correlation among values in the
observer frame.  If that fails, as it does for the {\it Swift}~sample, 
we can argue against (2) by
attempting to show that the source frame correlation represents
a tighter clustering.  The source frame clustering then presumably leads to
the observer frame clustering.  To demonstrate a significantly tighter clustering in the source frame, we
must control for the increase in clustering which arises due to multiplication by factors
containing redshift.

If we separate the terms in the $E_{\rm iso}$-$E_{\rm pk}$ powerlaw fit containing redshift from
those not containing redshift \citep[e.g.,][]{np05}, we see that the redshift independent terms
$\approx S^{0.5}_{\rm bol}/E_{\rm pk,obs}$ ($\approx \sqrt{n_{\rm bol}/E_{\rm pk,obs}}$)
exhibit a narrow scatter of $0.32\pm0.05$ dex.  (There is a strong observer frame 
correlation between our best-fit $S_{\rm bol}$ and $E_{\rm pk,obs}$: $\tau_K=0.49$, $5.7\sigma$.)
This is consistent with the scatter about the $E_{\rm iso}$-$E_{\rm pk}$
relation (Section \ref{sec:discuss}).
Stated differently, we can ignore the actual data and randomly sample the ratio $S_{\rm bol}/E_{\rm pk,obs}$
from a lognormal distribution and recover an equally significant (but fake)
$1+z$ versus $E_{\rm iso}/E_{\rm pk}$ correlation in a majority (94\%) of simulations.
The narrow scatter in these observer frame quantities is likely a consequence of the detector threshold, 
because there is less scatter
in $\sqrt{n_{\rm bol}}$ alone ($0.23\pm 0.04$ dex) and in $\sqrt{n_{\rm bol}/T^{0.5}_{90}}$ ($0.19\pm0.03$ dex;
Figure \ref{fig:thresh}; Top Left).
Moreover, the scatter in $\sqrt{n_{\rm bol}/T^{0.5}_{90}}$ ($0.24\pm0.04$ dex) changes little if we 
also include the short-duration and underluminous events.  

In addition to a weak dependence on the actual fluence and $E_{\rm pk,obs}$ data, little 
source frame information appears to be encoded in the $E_{\rm iso}$-$E_{\rm pk}$
correlation.  If we shuffle the redshifts among the sample, drawing
with repetition one redshift for each burst and recalculating the
source frame quantities given the observer frame quantities, we find
that a stronger correlation happens by chance in a large fraction 
(42\%) of simulations.  \citet{shaf07} discuss in detail the redshift degeneracy in
the $E_{\rm iso}$-$E_{\rm pk}$ and other high-energy relations.

We show now how the $E_{\rm iso}$-$E_{\rm pk}$ correlation can arise due to partial correlation
with the detector threshold.  First we divide the $E_{\rm iso}$, $E_{\rm pk}$
through by $E_{\rm pk,obs}$ and find that $(1+z)$ and $E_{\rm iso}/E_{\rm pk,obs}$ are tightly
correlated ($\tau_K=0.53$, $6.2\sigma$ significance).  The expected threshold for $E_{\rm iso}/E_{\rm pk,obs}$ scales
as $D_L^2/(1+z)\sqrt{T_{90}}$.  Controlling for partial correlation with this
quantity \citep[see,][]{as96}, we find that the $E_{\rm pk}$-$E_{\rm iso}$ correlation 
has $\tau_{K,{\rm partial}}=0.12$, with a significance of only $1.5\sigma$.
We conclude that the detector threshold accounts for a substantial portion of the $E_{\rm iso}$-$E_{\rm pk}$ correlation
seen for {\it Swift}~events.

Previous studies \citep[e.g.,][]{amati02} have concluded that the $E_{\rm pk}$--$E_{\rm iso}$ correlation is not due to
a flux selection
effect because individual burst fluences tend to lie well above the detector threshold.  This is not a strong enough
argument, however,
because even highly significant detections can be affected by a threshold bias if the values follow the threshold and are
clustered (e.g., Figure \ref{fig:thresh}; Top Left).  We can then invent apparent source
frame correlations
of arbitrarily small significance (i.e., very significant) by multiplying by steeper and steeper function of redshift.

It is also argued that the large dynamic range in $E_{\rm pk,obs}$ and fluence over which the
pre-{\it Swift}~$E_{\rm pk}$--$E_{\rm iso}$ relation is observed makes it less likely to be a selection effect.
However, the ratio of fluence to $E_{\rm pk,obs}$ is narrowly-distributed (i.e., these quantities are highly
correlated in the observer frame; Figure \ref{fig:catalog}, Top Right), even for the pre-{\it Swift}~sample \citep[see, e.g.,][]{amati02,taka05}, 
and this ratio is what we should require to exhibit a large dynamic range.

\subsection{Other High-Energy Correlations}
\label{sec:other}

A number of correlations have been reported in the literature among GRB timing
and spectral parameters in addition to the $E_{\rm iso}$--$E_{\rm pk}$ 
correlation.  We can test three of these against the {\it Swift}~data set
in Tables 1 and 2.

\citet{firmani06} have performed a careful search over high-energy parameters in
fits to 27 pre-{\it Swift}~GRBs
in order to find a very tight correlation $L_{\rm iso}
\approx 10^{48.5\pm0.1} ( E_{\rm pk}^{1.62} T^{-0.49}_{z,r45} )$.
We find a consistent
but less tightly constrained relation from the {\it Swift}~sample:
$L_{\rm iso} \approx 10^{48.5\pm0.7} ( E_{\rm pk}^{1.62} T^{-0.49}_{z,r45} )^{0.86\pm0.15}$.
Here, the duration $T_{r45}$ (Tables 1 \& 2) is transformed to the source frame according
to the prescription in \citet{firmani06}, 
$T_{z,r45}r = T_{r45}/(1+z) ( (1+z)/2 )^{0.4}$,
which accounts for the BAT energy range.
The most-probable new relation has large intrinsic scatter of $0.58\pm 0.8$ dex.

The \citet{firmani06} relation broadens in the {\it Swift}~sample due to the excess of hard and low-luminosity
events.  We find that most {\it Swift}~events (60\% or $38/63$) are inconsistent with the
pre-{\it Swift}~\citet{firmani06} relation at the 90\% confidence level.
Stated differently, most (71\%) of the 
{\it Swift}~GRB 90\% confidence redshift intervals
on a redshift estimator $\hat z$
determined assuming the relation do not contain the actual redshift.  

Using Equation \ref{eq:scatx} to account for the large uncertainties in $\log_{10}{(1+\hat z)}$,
we find that there is a large intrinsic scatter of $0.6\pm0.1$ dex between this and the same
of function using the host galaxy/spectroscopic redshift.  We find that
50\% of the probability in terms of estimated redshift for 40\% of
events is at $z>10$,
due to the faintness and hardness of the {\it Swift}~events.
Ignoring these cases, the 90\% confidence redshift estimate
still fails for 53\% ($20/38$) of events.  There is
only a weak correlation ($\tau_K=0.17$, $1.9\sigma$) between the best-fit redshift assuming
the relation and the actual redshift.

The {\it Swift}~\citet{firmani06} correlation, using the best-fit parameters from Tables 1 \& 2, has $\tau_K=0.61$ ($7.1\sigma$). However,
this decreases to $\tau_{K,{\rm partial}} = 0.13$ ($1.5\sigma$) if we control for partial correlation
with the detector threshold.
Because the ratio of source frame quantities used in the \citet{firmani06} relation is narrowly-distributed for the
{\it Swift}~sample (0.45 dex), we find that most simulations (71\%) of the source frame correlation using fake observer frame
data yield a more significant source frame correlation.

We observe a significant correlation between $L_{\rm iso}$ and $E_{\rm peak}$
\citep[see,][]{shaf03,yon04}, with $\tau_K=0.55$ ($6.3\sigma$) for
our best-fit model parameters (Tables 1 \& 2).  However,
if we control for partial correlation with the detector threshold as above,
the correlation largely disappears ($\tau_{K,{\rm partial}}=0.15$, $\sigma=1.8$).
Most simulations (96\%) of the source frame correlation using fake observer frame
data yield a more significant source frame correlation.

\citet{atteia03} find a tight correlation between $N_{\rm iso}/E_{\rm pk,obs}/\sqrt{T_{90}}$
and $1+z$ for data detected by the {\it HETE-2}~satellite, which is used as a redshift
estimator.  We find a modestly significant correlation ($\tau_K=0.38$, $4.4\sigma$)
among our best fit parameters (Tables 1 \& 2).
As hinted by the similar form of the \citet{atteia03} redshift estimator to our
$S/N$ estimator ($n_{\rm bol}/\sqrt{T_{90}}$; Section \ref{sec:thresh}), the correlation strength
degrades greatly when we control for partial correlation with the detector threshold
($\tau_{K,{\rm partial}}=0.14$, $1.6\sigma$).  
Most simulations ($>$99\%) of the source frame correlation using fake observer frame
data yield a more significant source frame correlation.

We note that the \citet{atteia03} correlation and the correlations above can be used as redshift
estimators only if one believes that the correlation exists in the source frame, independent
of the observer frame detector threshold.

We do not test here the veracity of correlations between temporal lag 
and luminosity \citep[e.g.,][]{nmb00} or temporal variability and luminosity \citep[e.g.,][]{fr00}. However, we stress
that future analyses testing these correlations must account for the detector
threshold.

We also do not test the correlation between $E_{\rm pk}$ and the beaming corrected 
energy release $E_{\gamma}$ found by \citep{ggl04} \citep[see, also,][]{liangzang05}. 
This requires measurement of
late-time light curve breaks, and these appear to be ambiguous
or non-existent for many {\it Swift}~events \citep{sato07,willin07}.
\citet{pain07} find that the $E_{\rm pk}$-$E_{\gamma}$ exists in the {\it Swift}~sample
but is largely a consequence of the $E_{\rm pk}$-$E_{\rm iso}$ correlation.  We explore
this issue in a separate paper \citep{dans_paper}.

\subsection{The pre-{\it Swift}~Threshold and Low-$z$ Events}
\label{sec:lowz}

We assume here that the pre-{\it Swift}~correlations represent measures of detection
threshold.  In addition to the instrument that detected the GRB (e.g., {\it HETE-2}~or {\it Beppo-SAX}),
this includes biases due to the source localization, optical afterglow brightness and
host galaxy brightness and star formation rate.  There are also likely strong biases due to outlier
rejection during the construction of the correlations.  The \citet{firmani06} relation
is potentially the best measure of the sum total of these biases, because that relation has the
narrowest scatter.  

In Figure \ref{fig:firmani_thresh}, we plot the observer-frame quantities used to form the \citet{firmani06}
relation versus redshift.  There is a moderately strong correlation in the \citet{firmani06} sample ($\tau_K=0.76$, $4.6\sigma$).
Excluding the underluminous and short-duration {\it Swift}~events as above, the {\it Swift}~sample exhibits lower flux values, and the correlation 
among best-fit model parameters (Table 2) is weak ($\tau_K=0.17$, $1.9\sigma$).  Because the \citet{firmani06}
relation has a similar form to our detector threshold estimate (Section \ref{sec:thresh}), as we describe below, the
clustering in observer-frame quantities in Figure \ref{fig:firmani_thresh} is likely a consequence of a clustering above threshold
(Figure \ref{fig:thresh}; Top Left).  

Additionally, it is plausible that the moderately strong correlation between these quantities and redshift
in the pre-{\it Swift}~sample arises from biases associated with optical transient detection.  
If the intrinsic optical flux tracks the $\gamma$-ray flux, then there will be a truncation below lines roughly parallel
to the dotted line in Figure \ref{fig:firmani_thresh}.
We find that bright ($R<18$ mag) optical
transients were not detected within 0.5 days of the GRB for 25\%
of events above $5 \times 10^{-10}$ erg cm$^{-2}$ s$^{-0.51}$ keV$^{-1.62}$ in Figure \ref{fig:firmani_thresh}.  However, more than half of events below a line drawn
at that value do not have bright optical counterparts.  Pre-{\it Swift}~host galaxy associations depended on tight localizations
possible only from optical detections, whereas {\it Swift}-era host galaxies are often pin-pointed from XRT observations.

The redshift dependence of the observer frame quantities in the pre-{\it Swift}~sample appears compelling in large part due to the
presence of the bright GRB~0303029 (Figure \ref{fig:firmani_thresh}) at low redshift.  However, the {\it Swift}~sample shows the
high flux of this event to be anomalous.  There are additional points --- overlapping with the {\it Swift}~cyan points
in Figure \ref{fig:firmani_thresh} --- which were discarded as outliers by \citet{firmani06}.

If we~re-write the \citet{firmani06} relation in terms of $n_{\rm bol}$,
we find 
\begin{multline}
  n_{\rm bol,firmani} \approx 30 \left \{{ E_{\rm pk,obs} \over 100~{\rm keV} } \right \}^{0.62} T_{r45}^{-0.49} (1+z)^{-0.09} \\
  \times \left \{ {D_L(z=1) \over D_L } \right \}^2 \left \{ {1+z \over 2} \right \}^2 \gamma ~{\rm cm}^{-2}.
    \label{eq:firmani}
\end{multline}
Equation \ref{eq:firmani} has a similar time dependence to our threshold.  
There is also a weak
dependence on the burst $E_{\rm pk,obs}$.  
Ignoring the difference between $T_{90}$ and $T_{r45}$, 
the value of the threshold is in an order of magnitude lower than our 
estimated {\it Swift}~threshold (Section \ref{sec:thresh}).
This is consistent with
the relative fraction of expected hard/faint events in the {\it Swift}~and pre-{\it Swift}~samples
determined in Section \ref{sec:discuss}.  The redshift variation in Equation
\ref{eq:firmani} 
 is extremely weak for $z\gtrsim 1$.

An explanation in terms of a threshold provides a natural explanation for why low-$z$ events like GRBs 051109B, 060218, 060614,
and the short-duration GRBs do not fit on the \citet{firmani06} and other 
relations.  In Sections \ref{sec:thresh} and \ref{sec:other}, we discuss
how these {\it Swift}~short-duration and underluminous events, which are strong outliers to the 
$E_{\rm iso}$--$E_{\rm pk}$ relation, do follow the source frame correlation set by the detector threshold (Figure \ref{fig:thresh}; Bottom Left).

A test of our claim that most pre-{\it Swift}~high-energy correlations are due simply to the detector threshold is
to also show that the pre-{\it Swift}~underluminous and short duration events satisfy Equation (\ref{eq:firmani}).
In terms of $E_{\rm iso}$, if the events are near threshold, the expected value is 
\begin{multline}
  E_{\rm iso, thresh} \approx 5 \times 10^{49} \left \{{ E_{\rm pk,obs} \over 100~{\rm keV} } \right \}^{1.62} \left \{{T_{r45} \over 5~{\rm s}}\right \}^{-0.49}  \\
  \times
    \left \{ {D_L \over D_L(z=0.1) } \right \}^2 \left \{ {1.1 \over 1+z} \right \}^2 ~{\rm erg}.
    \label{eq:firmani_thresh}
\end{multline}
The observed $E_{\rm iso}$ and $E_{\rm obs,pk}$ for GRB~980425 at $z=0.0085$ \citep[e.g.,][]{kouv04,ggl04} are $\approx 10^{48}$ erg and 120 keV, respectively.
If we assume $T_{r45}=5$ s, then we find $E_{\rm iso, thresh} \approx 5 \times 10^{47}$ erg.
The observed $E_{\rm iso}$ and $E_{\rm obs,pk}$ for GRB~031203 at $z=0.105$ \citep[e.g.,][]{saz04} are $\approx 5 \times 10^{49}$ erg and $\gtrsim 190$ keV, respectively.
If we assume $T_{r45}=5$ s, then we find $E_{\rm iso, thresh} \approx 5 \times 10^{49}$ erg.
The observed $E_{\rm iso}$ and $E_{\rm obs,pk}$ for short GRB~050709 at $z=0.16$ \citep[e.g.,][]{joel05} are $\approx 10^{50}$ erg and 84 keV, respectively.
If we assume $T_{r45}=0.5$ s, then we find $E_{\rm iso, thresh} \approx 10^{50}$ erg.  This agreement is excellent.

Because we can understand the luminosities of these events in terms of a detector threshold, there is little reason
to think of them as anomalously subluminous \citep[see, e.g.,][]{sod04,guetta04,watson06,ghis06}.
If our threshold versus redshift is in fact correct for short durations, then there is little reason to believe
that the intrinsic energy release in short durations GRBs is different from that in long duration GRBs.

\section{Conclusions}

\begin{figure}
\includegraphics[width=3.5in]{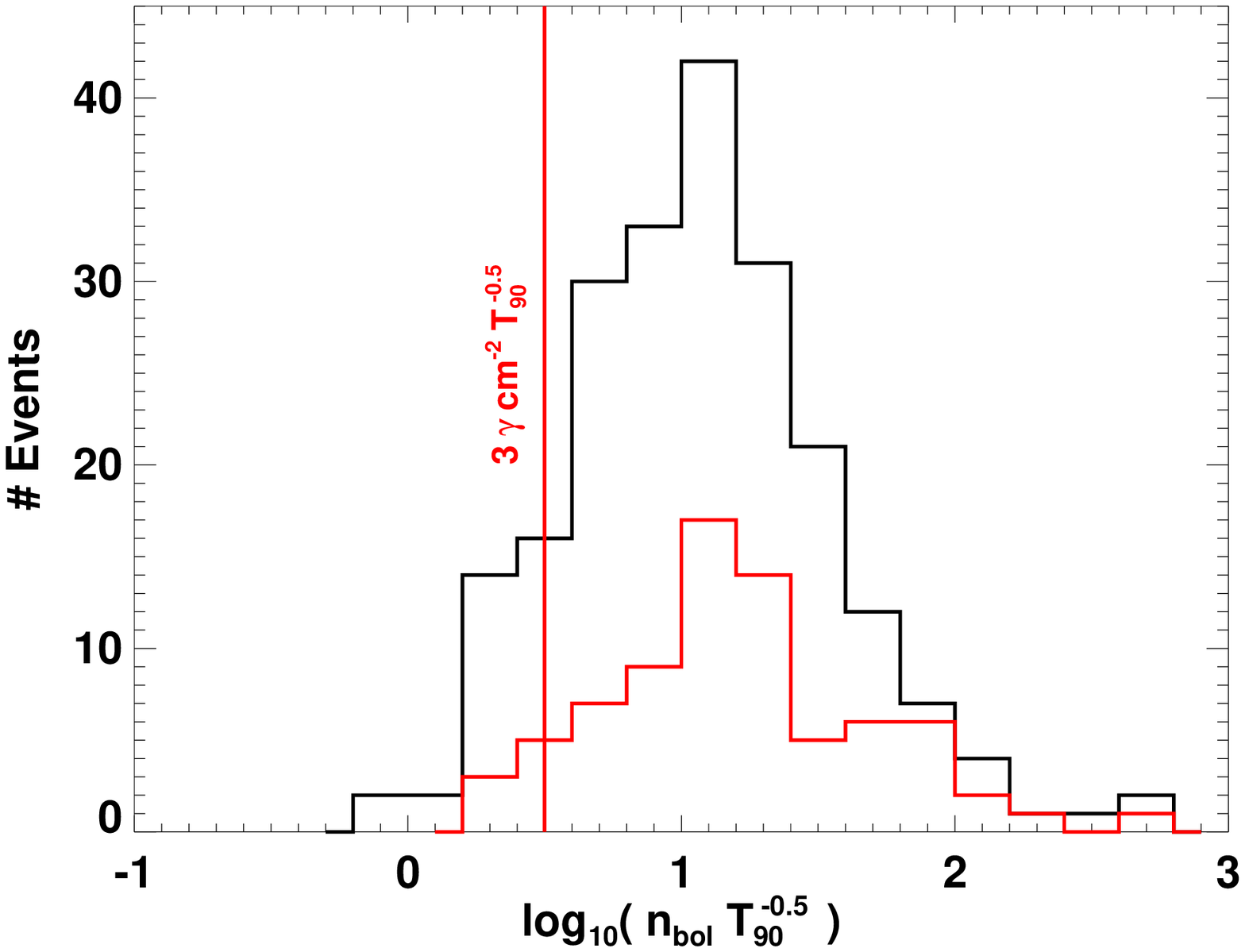}\includegraphics[width=3.5in]{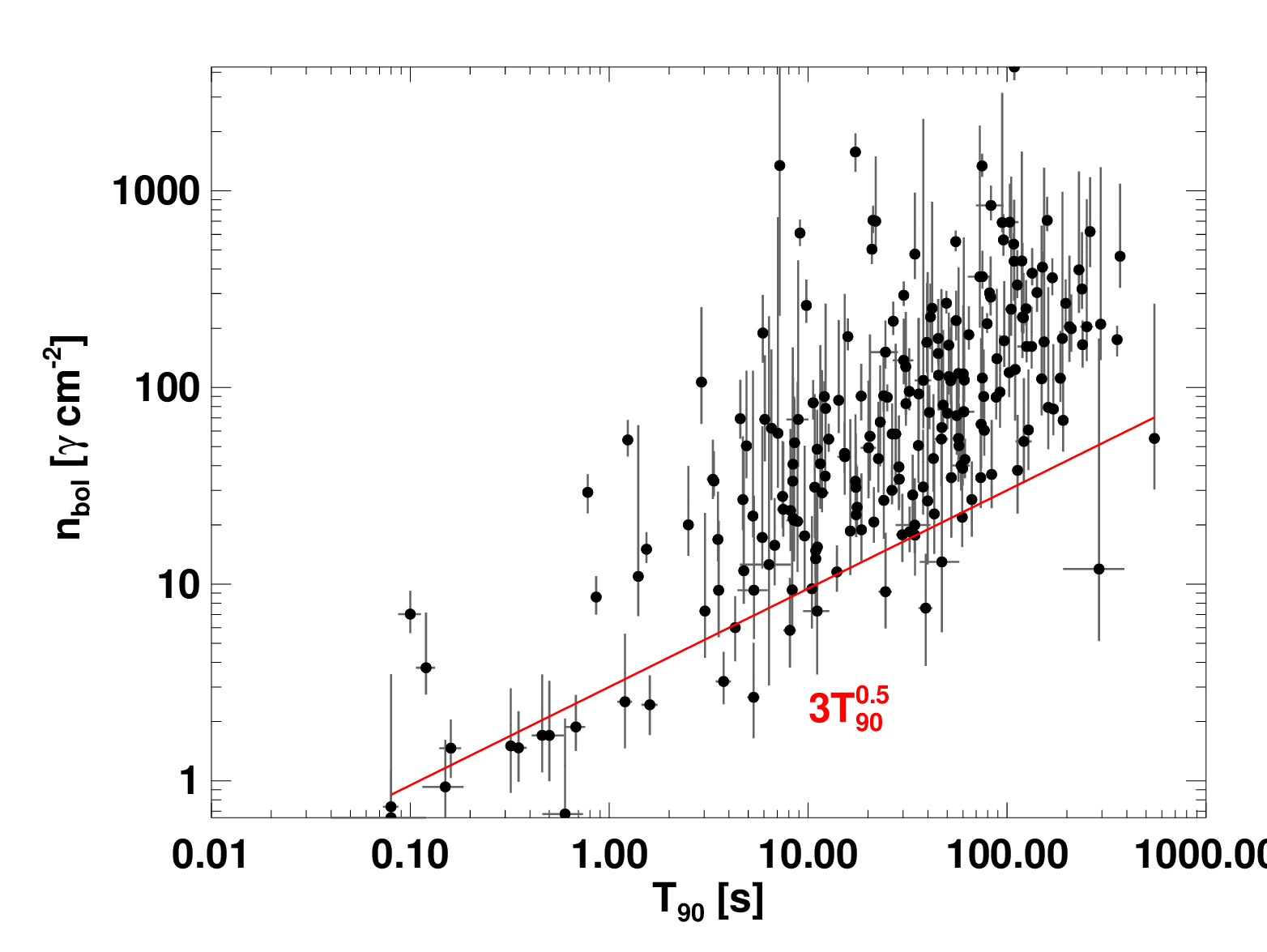}
\includegraphics[width=3.5in]{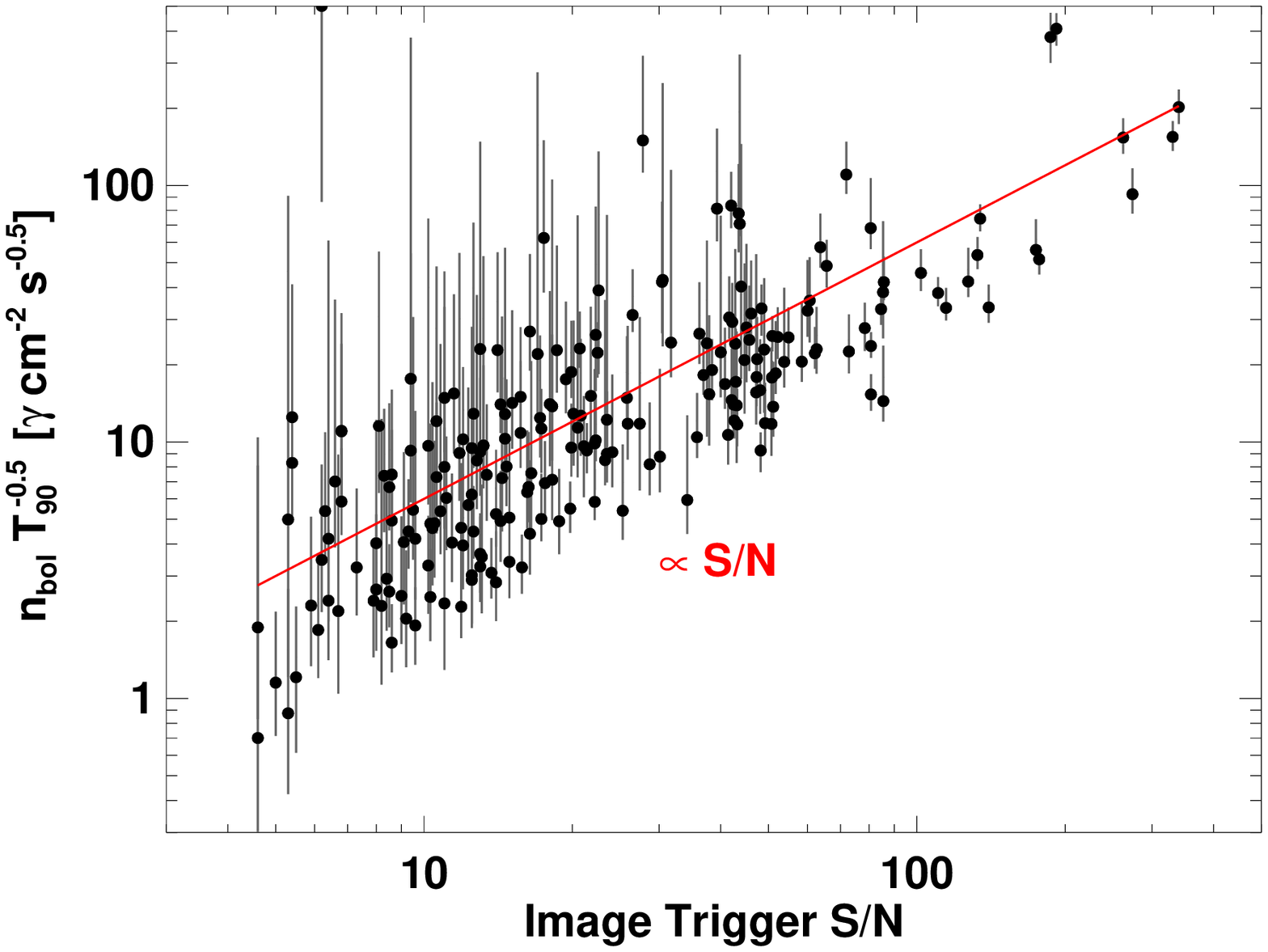}\includegraphics[width=3.5in]{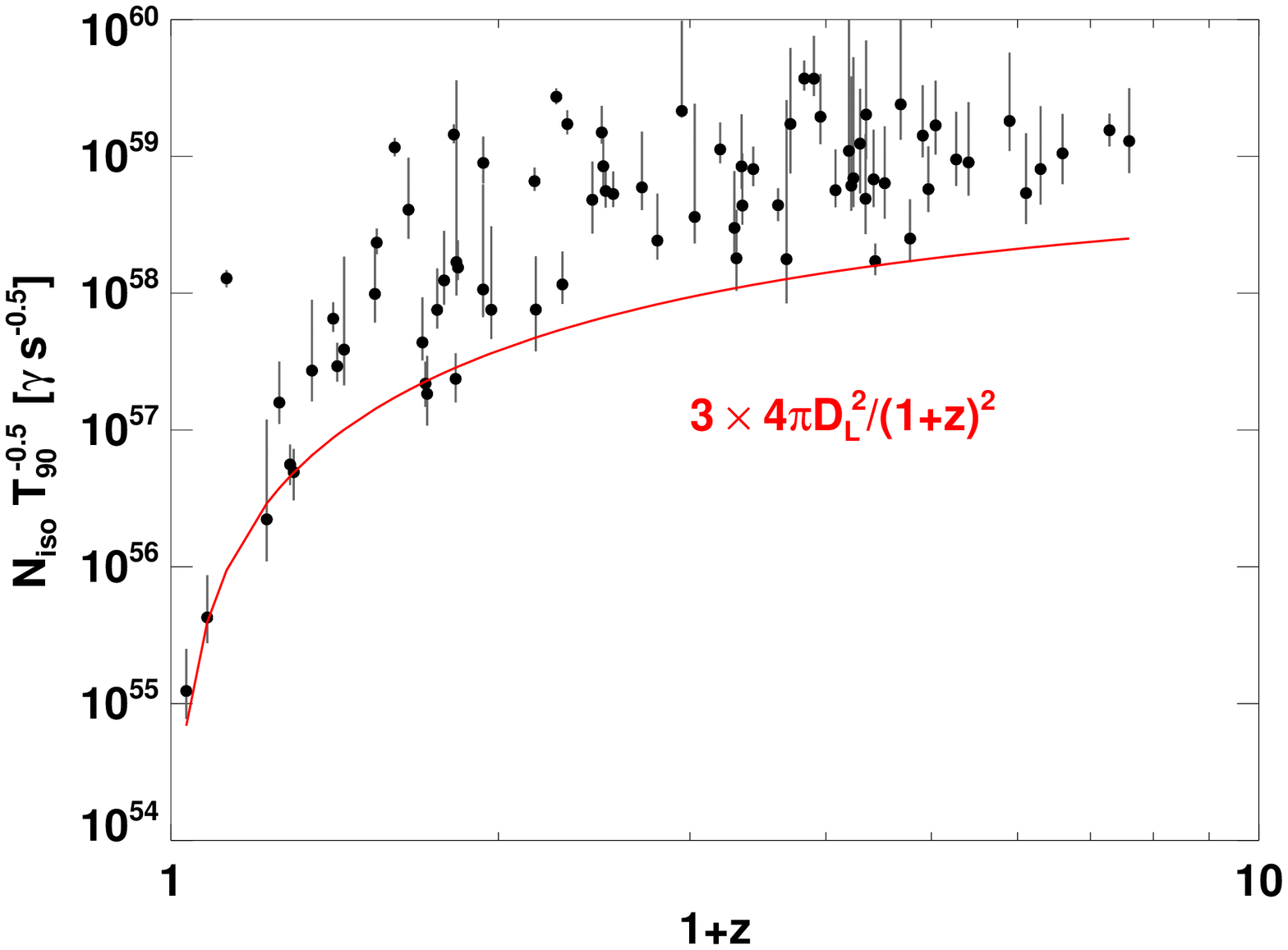}
\caption{\small Experimental determination of the BAT threshold in terms of $n_{\rm bol}/\sqrt{T_{90}}$.  (Top Left) The
histogram of 1--$10^4$ keV photon fluence $n_{\rm bol}$ (Table 2) divided by the 
square root of $T_{90}$ (Table 1) exhibits a narrow
distribution and 90\% of values are greater than a threshold of 3 ph cm$^{-2}$ s$^{-0.5}$.
The red histogram shows the events with measured redshift.
(Top Right) Scatter plot showing a correlation between $n_{\rm bol}$ and $T_{90}$ likely due to the threshold.
The approximate threshold from the Top Left plot is also plotted.  (Bottom Left) The trigger $S/N$ (Table 1) correlates
tightly with $n_{\rm bol}/\sqrt{T_{90}}$.  The best fit powerlaw (plotted) has index $\approx 1$ and 
normalization $S/N\approx 5$ at $n_{\rm bol}/\sqrt{T_{90}}
= 3$ ph cm$^{-2}$ s$^{-0.5}$.  (Bottom Right) The effect of the threshold in the source frame versus redshift $z$.}
\label{fig:thresh}
\end{figure}

\begin{figure}
\centerline{\includegraphics[width=4.5in]{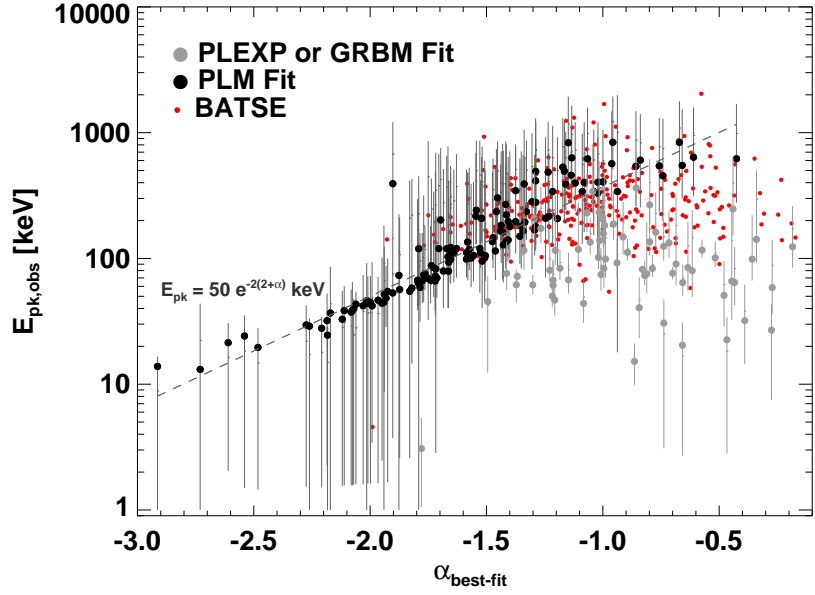}}
\caption{The role of the Bayesian priors over a large $E_{\rm pk,obs}$ range plotted versus best-fit powerlaw photon
index $\alpha_{\rm best-fit}$.  The dark black points and error bars show $E_{\rm pk,obs}$
determinations for events adequately fit by a powerlaw model (PLM).  In the classical frequentist approach to
spectral fitting, these spectra yield only limits (Table 2).  Very large $E_{\rm pk,obs}$ values are
extremeley uncommon, as is clear from the red points from {\it BATSE}~\citep{kaneko06}, which are plotted
using the GRBM high energy index $\alpha$ on the x-axis;
we truncate the probability at high $E_{\rm pk,obs}$
through our model priors.  We do not truncate the probability at low $E_{\rm pk,obs}$ until 1 keV, and it is clear
that {\it Swift}~measures $E_{\rm pk,obs}$ values (grey points) below the {\it BATSE}~bandpass.
A strong correlation between the best-fit photon index $\alpha_{\rm best-fit}$ for the black points --- which is not
necessarily $\alpha$ or $\beta$ in the GRBM due to the possible proximity of $E_{\rm pk,obs}$ to the bandpass  ---
and $E_{\rm pk,obs}$ results (Section \ref{sec:kw}).}
\label{fig:alpha_ep}
\end{figure}

\begin{figure}
\includegraphics[width=3.5in]{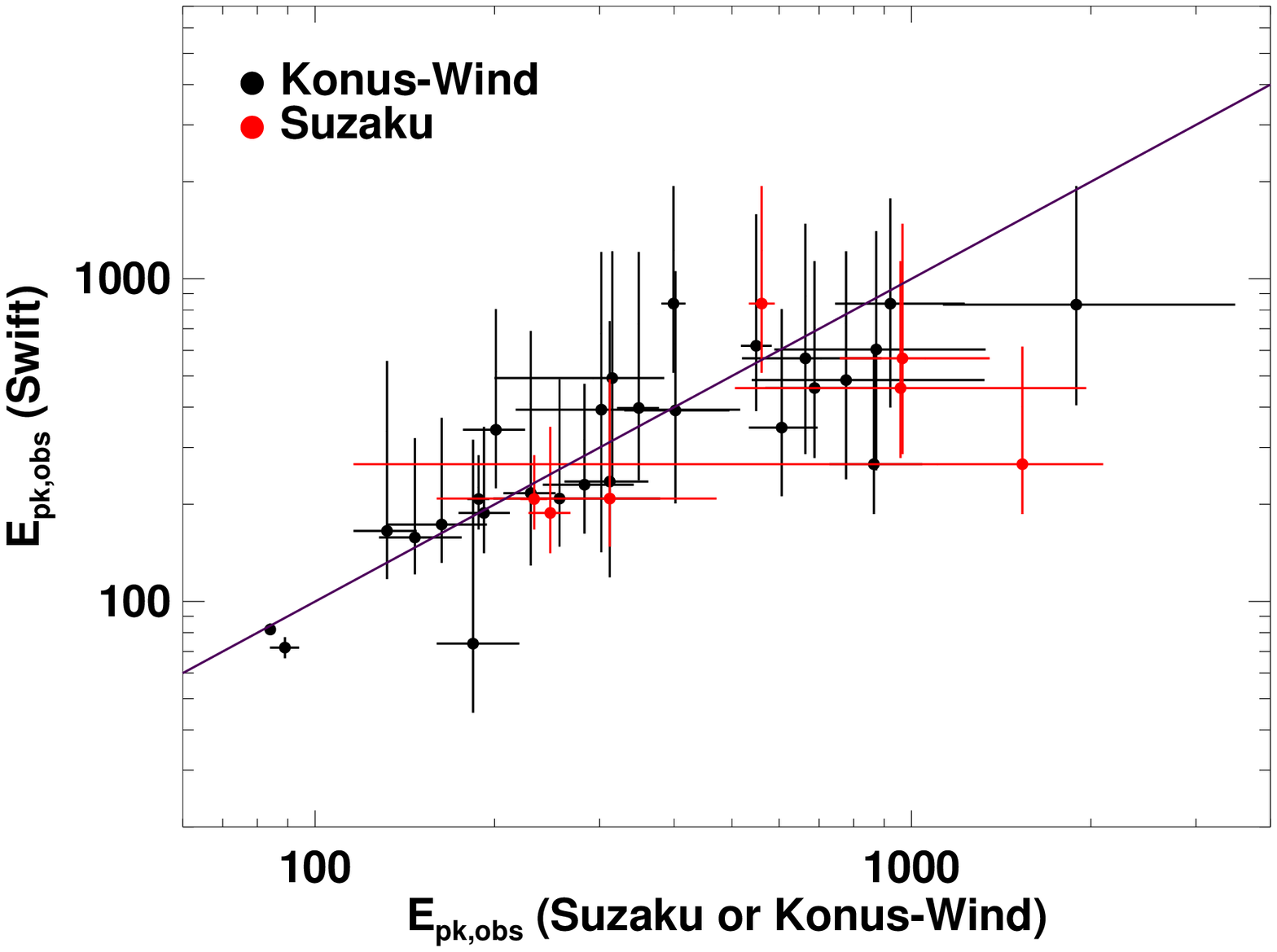}\includegraphics[width=3.5in]{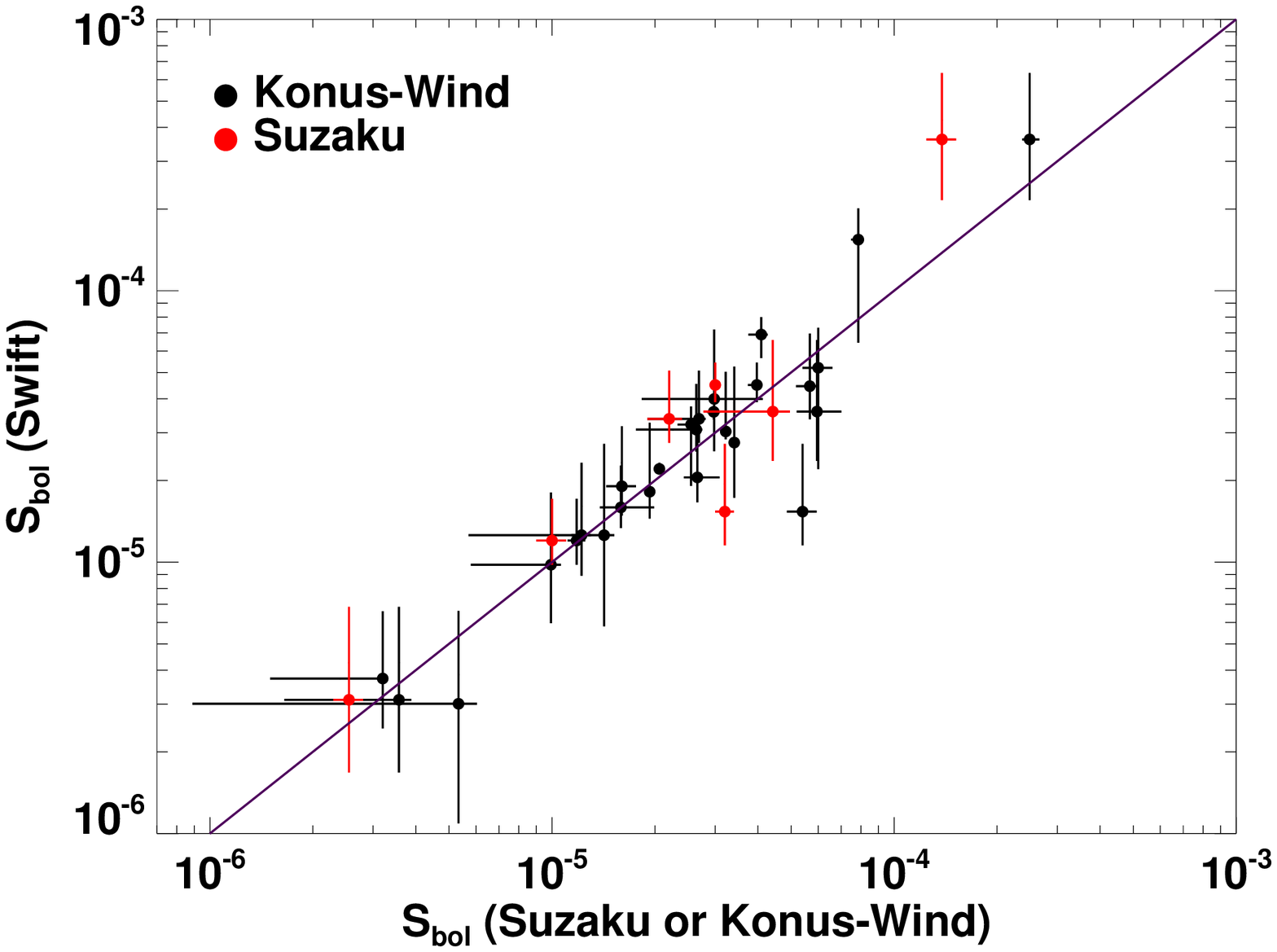}
\caption{\small Comparison of $E_{\rm pk,obs}$ and approximate bolometric fluences $S_{\rm bol}$ derived
here to values reported in the GCN circulars for Konus-Wind \citep[e.g.,][]{gol06} or Suzaku \citep[e.g.,][]{hong07} and the lines of equality.  Although
the BAT detector approaches zero effective area above 150 keV, we are able to accurately recover
the true $E_{\rm pk,obs}$ and fluence without bias.}
\label{fig:sks}
\end{figure}

\begin{figure}
\includegraphics[width=3.5in]{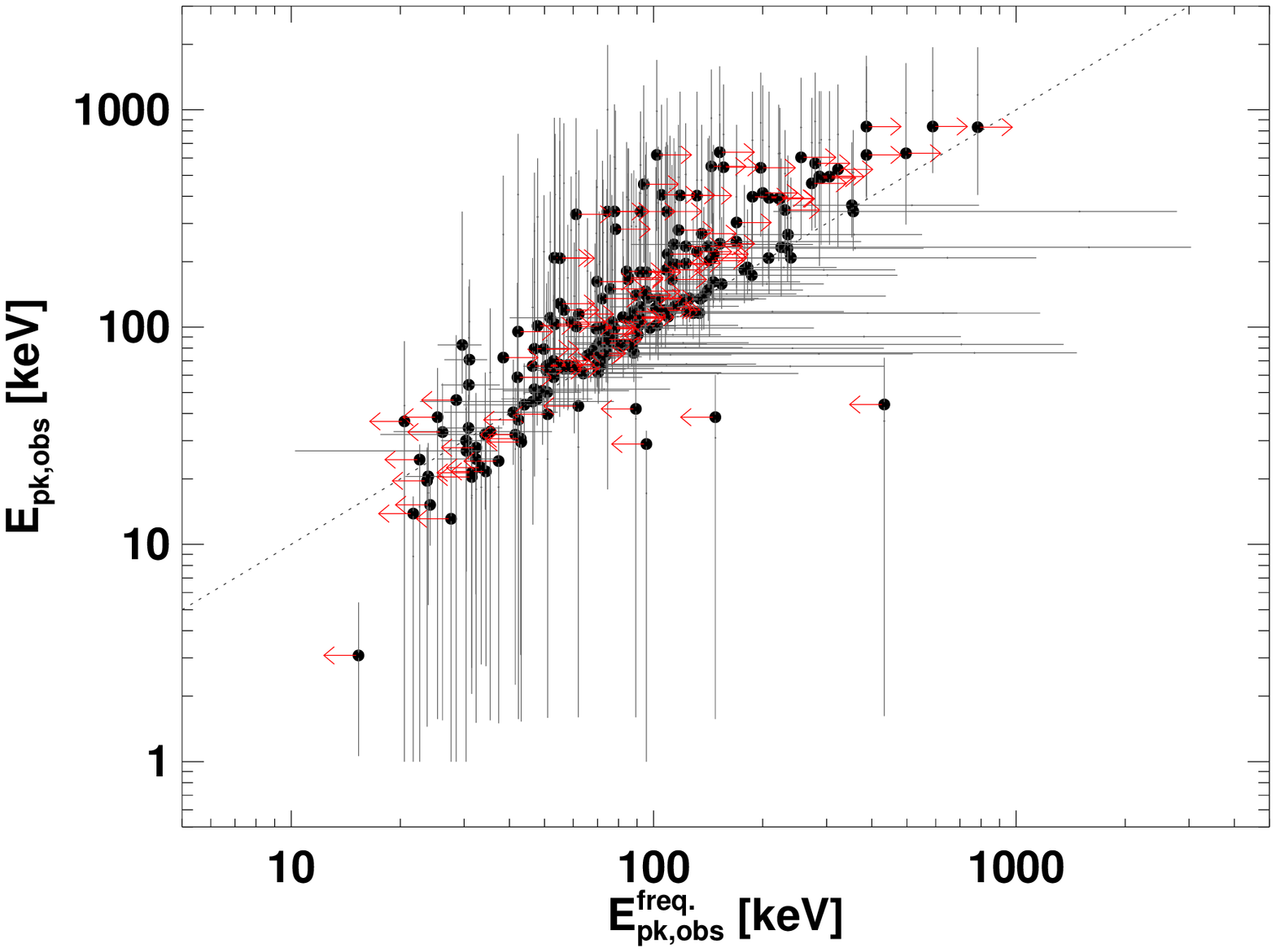}\includegraphics[width=3.5in]{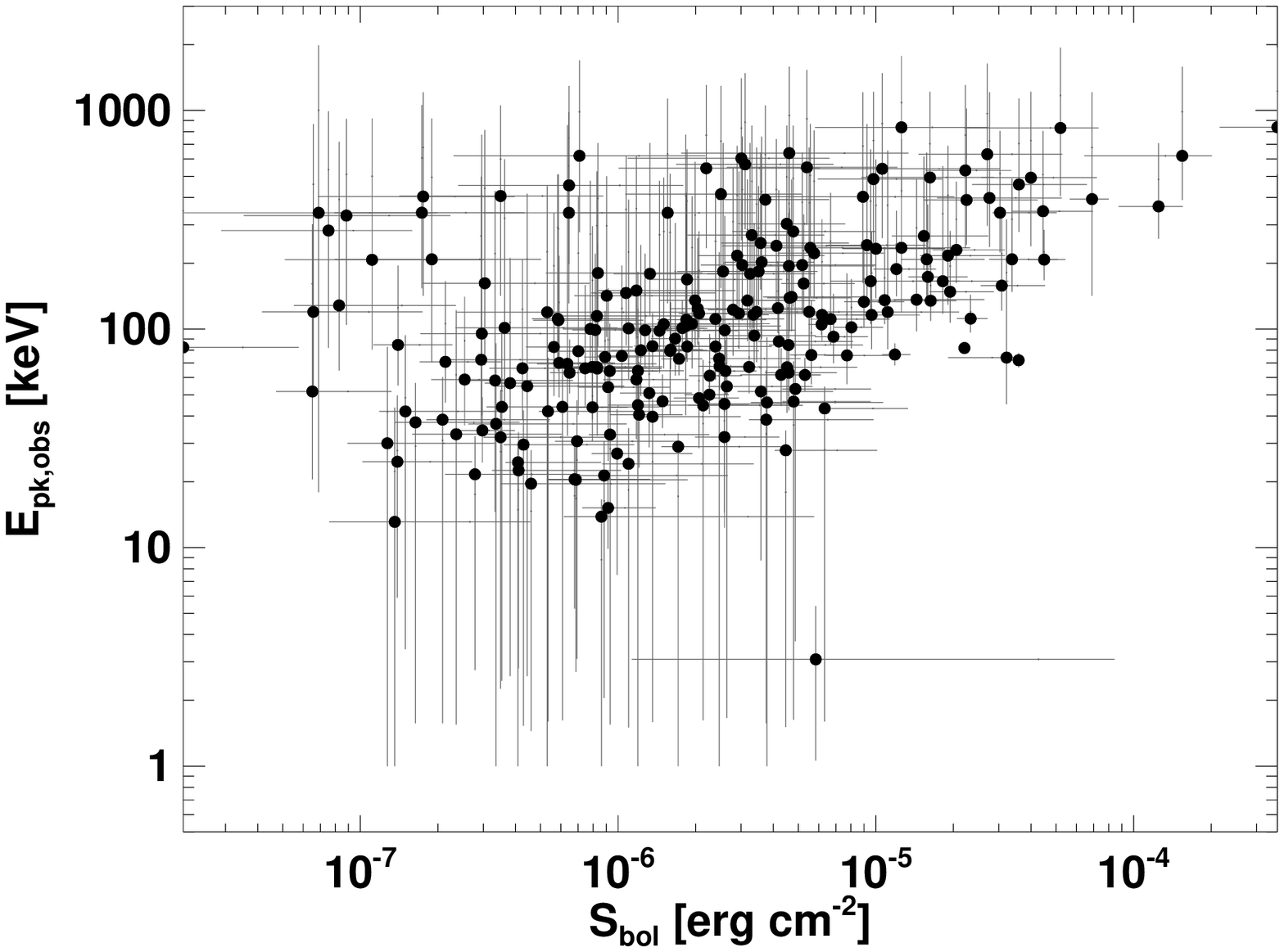}
\includegraphics[width=3.5in]{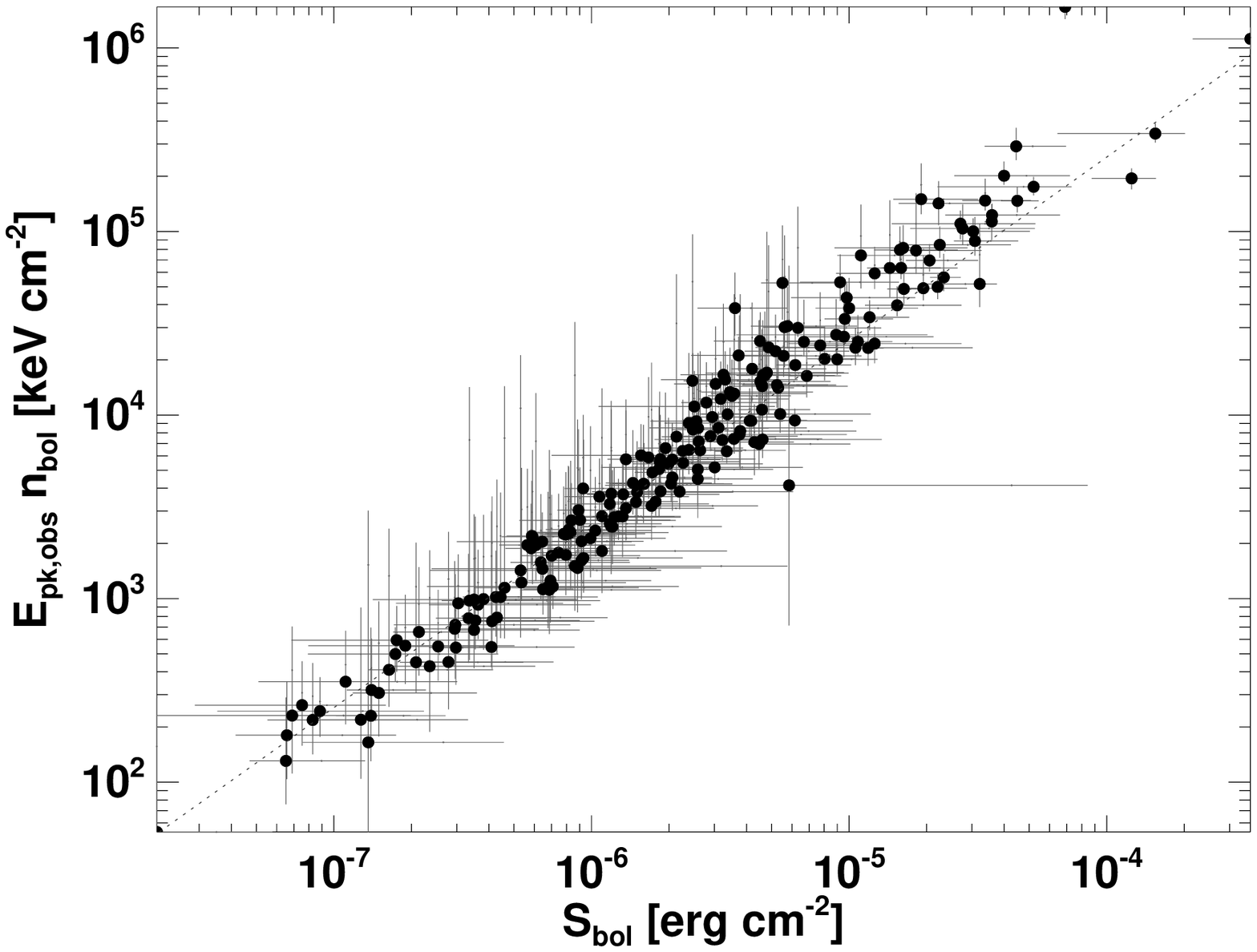}\includegraphics[width=3.5in]{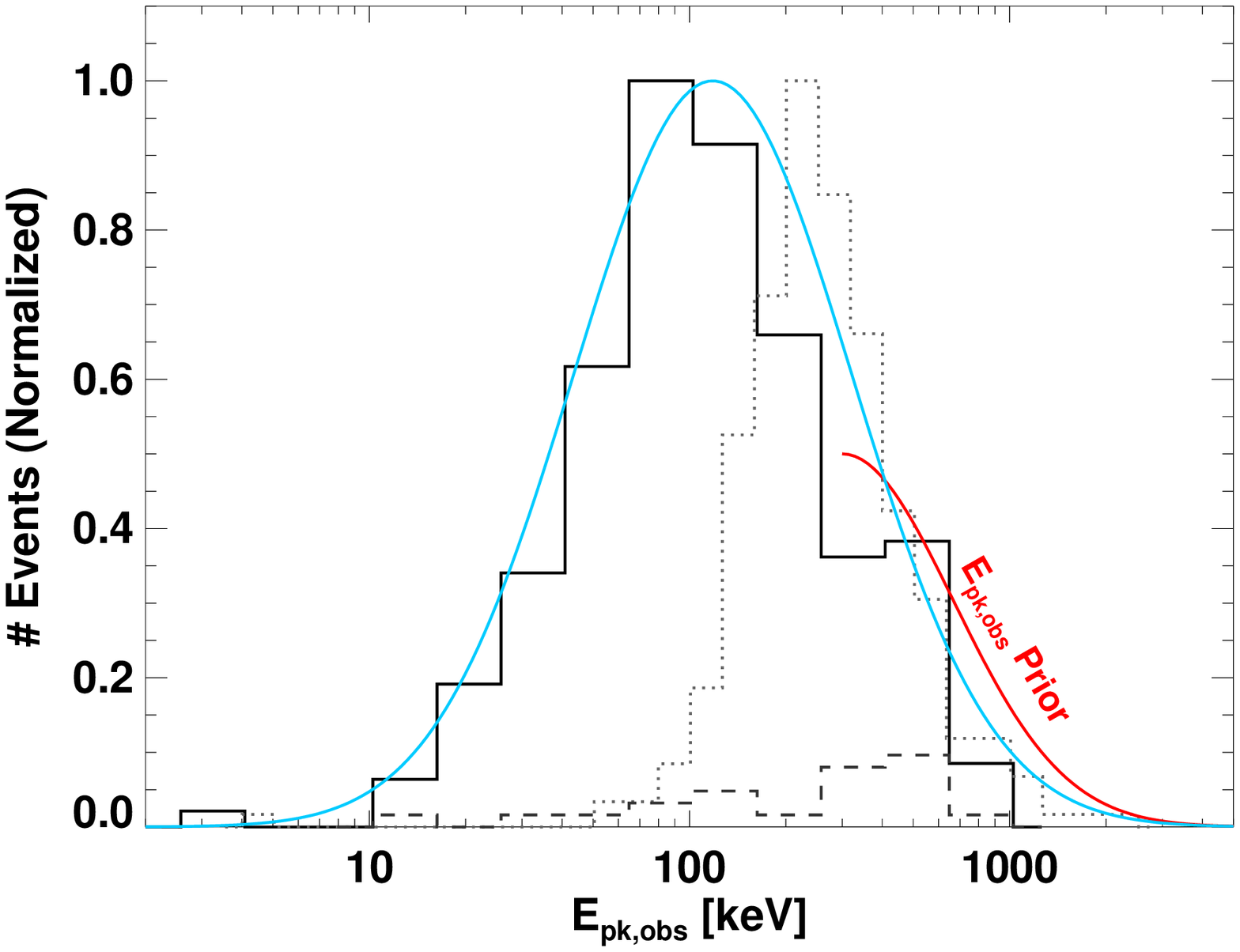}
\caption{\small Catalog values for the 1--$10^4$ keV photon ($n_{\rm bol}$) and energy
($S_{\rm bol}$) fluences and $\nu F_{\nu}$ spectral peak energy $E_{\rm pk,obs}$.
(Top Left) Bayesian versus frequentist $E_{\rm pk,obs}$ estimates and the line of equality.  The red
arrows designate $E_{\rm pk,obs}^{\rm freq.}$ limits for the majority (66\%) of BAT events.
(Top Right)  $E_{\rm pk,obs}$ versus the approximate bolometric fluence $S_{\rm bol}$.
(Bottom Left) The approximately bolometric photon fluence scales as $S_{\rm bol}/E_{\rm pk,obs}$.
(Bottom Right) The {\it Swift}~sample contains an excess of low $E_{\rm pk,obs}$ events, similar
to {\it HETE-2}~\citep{taka05}, compared to the
\citet{kaneko06} catalog of bright {\it BATSE}~GRBs.
In blue we plot the fit to the $E_{\rm pk,obs}$ (including errors) for each GRB (from Equations \ref{eq:onex} and \ref{eq:scatx}).  We assume the red line
as a prior in the fitting for each event in order to measure upper confidence interval bounds on $E_{\rm pk,obs}$.
The histogram for {\it Swift}~``short-duration'' GRBs is given as a dashed histogram.}
\label{fig:catalog}
\end{figure}

\begin{figure}
\centerline{\includegraphics[width=6.0in]{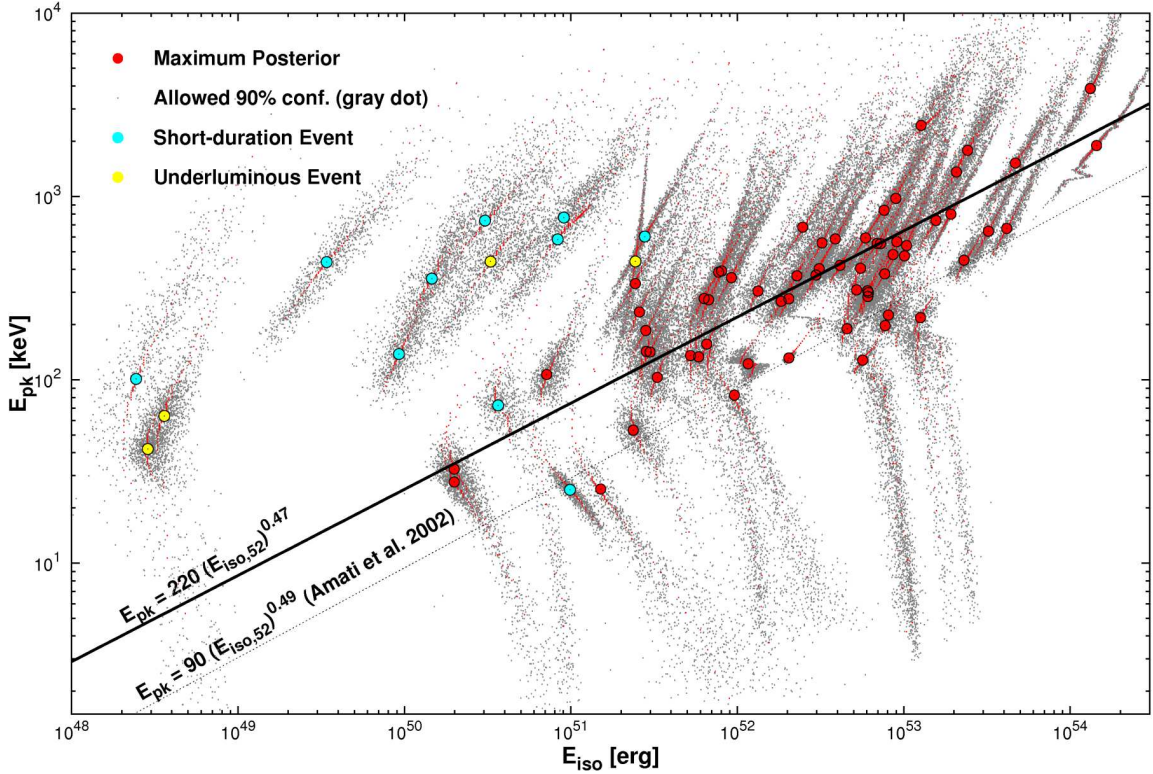}}
\caption{\small Maximum posterior probability values for $E_{\rm pk}$ and $E_{\rm iso}$ (filled circles)
for 77 {\it Swift}~GRBs with measured redshifts (Table 2).  Also plotted are samples from the 90\%
confidence region in $E_{\rm pk}$ and $E_{\rm iso}$ for each GRB (small gray dots).  Outliers in terms
of short burst duration or low-luminosity are
labelled.  The best fit lines from this study (excluding the underluminous and ``short-duration'' events)
and from the Beppo-SAX sample of \citet{amati02} are also plotted.  Events furthest from the \citet{amati02} line
in terms of low $E_{\rm iso}$ and high $E_{\rm pk}$ are closer to detector thresholds (Figure \ref{fig:amati_thresh}).}
\label{fig:amati}
\end{figure}

\begin{figure}
\centerline{\includegraphics[width=5.0in]{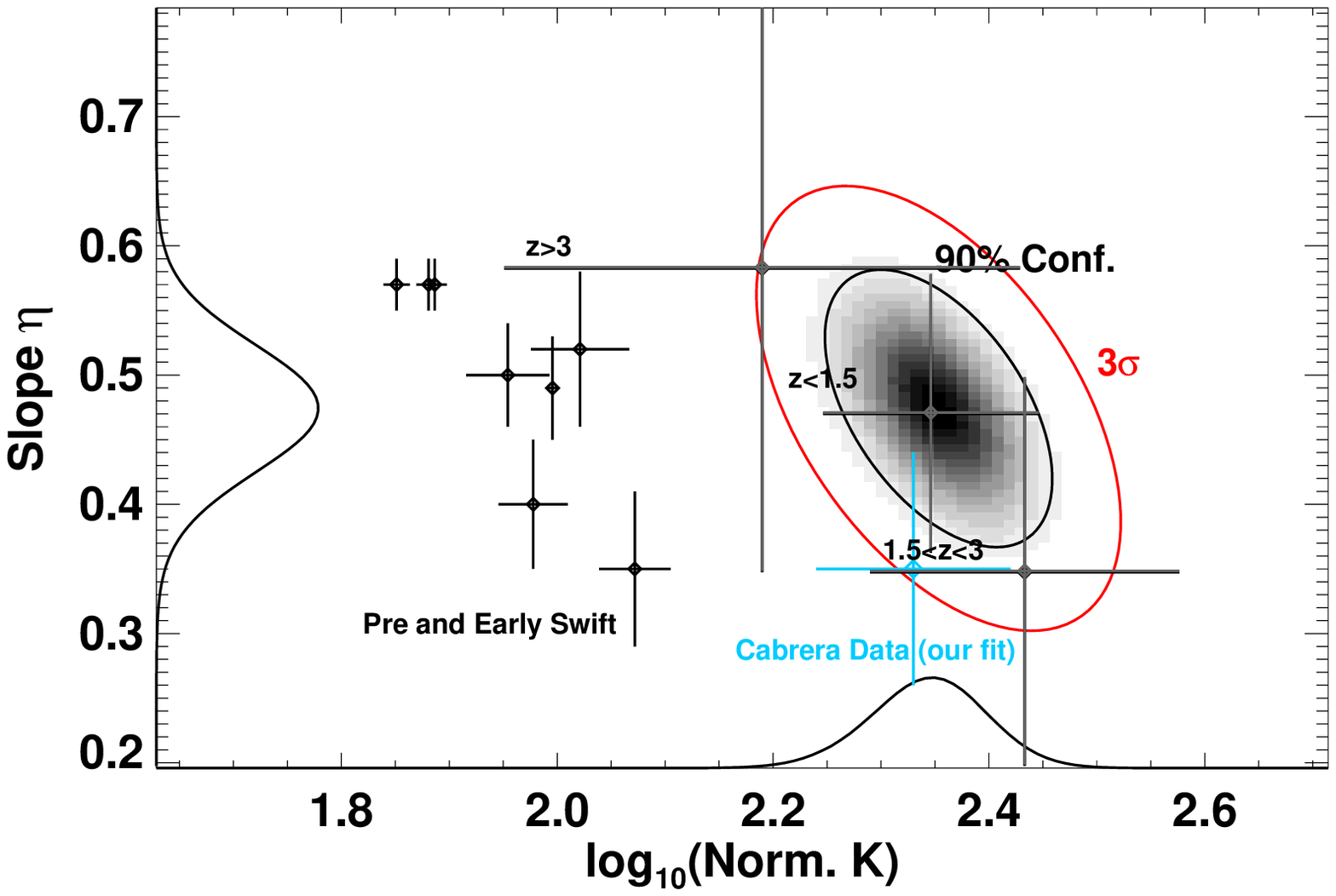}}
\caption{\small Posterior probability contours for a powerlaw relation $E_{\rm pk}=K~(E_{\rm iso}/[10^{52} {\rm ~erg}])^{\eta}$ keV.
We also plot ($K,\eta)$ for the sample divided into 3 redshift bins.
Estimates of the relation before {\it Swift}~and including early {\it Swift}~data summarized in \citet{amati06}
(and from \citet{amati02,amati03,ggl04,fb05,nava06})
are inconsistent with the current sample normalization $K$ at the $>5\sigma$ level.  $E_{\rm pk}$ and
$E_{\rm iso}$ values and confidence regions from \citet{cabrera07}
for 28 of the events also used in this study are also fit (cyan point; Section \ref{sec:cabrera}).
The marginal posterior
distributions in $K$ and $\eta$ are plotted above the $X$ and to the right of $Y$ axes, respectively.}
\label{fig:amati_prob}
\end{figure}

We fit for the durations and spectral parameters of 218 {\it Swift}~GRBs, including
77 GRBs with redshifts.  Unbiased estimates of $E_{\rm iso}$ and $E_{\rm pk,obs}$ 
(at least for $E_{\rm pk,obs}\lessim 1$ MeV)
are possible for these events if we adopt a Bayesian spectral fitting
approach, with relatively weak prior assumptions.  Because $E_{\rm pk,obs}$ is typically
poorly-constrained and correlates with $E_{\rm iso}$ in a complicated fashion, we rigorously propagate 
errors via a sampling of the posterior probability.  We have searched for correlations
among the observer frame quantities, and have found a family of correlations (Figure
\ref{fig:thresh}) which we argue to be due to the detector threshold.  We have also
measured apparently statistically significant correlations among host frame quantities
(Sections \ref{sec:amati_cor} and \ref{sec:other}).

Thanks to the large {\it Swift}~BAT dataset, we now understand the probable origin
of the $E_{\rm pk}$-$E_{\rm iso}$ correlation: a trigger threshold $\propto n_{\rm bol}$
and an intrinsic lack of bright relative to faint sources induces a strong 
observer-frame correlation between $E_{\rm pk,obs}$ and $S_{\rm bol}$.  The apparent correlation strength
is then amplified as we multiply both side of the equation by strong functions of
redshift, in order to transform to the host frame quantities.  The redshifts or fluence
and $E_{\rm pk,obs}$ data can be
drawn at random and do not have to correspond to the actual host galaxy redshifts
or measured values in order to produce a significant correlation in the host frame quantities.

\begin{figure}
\centerline{\includegraphics[width=6.0in]{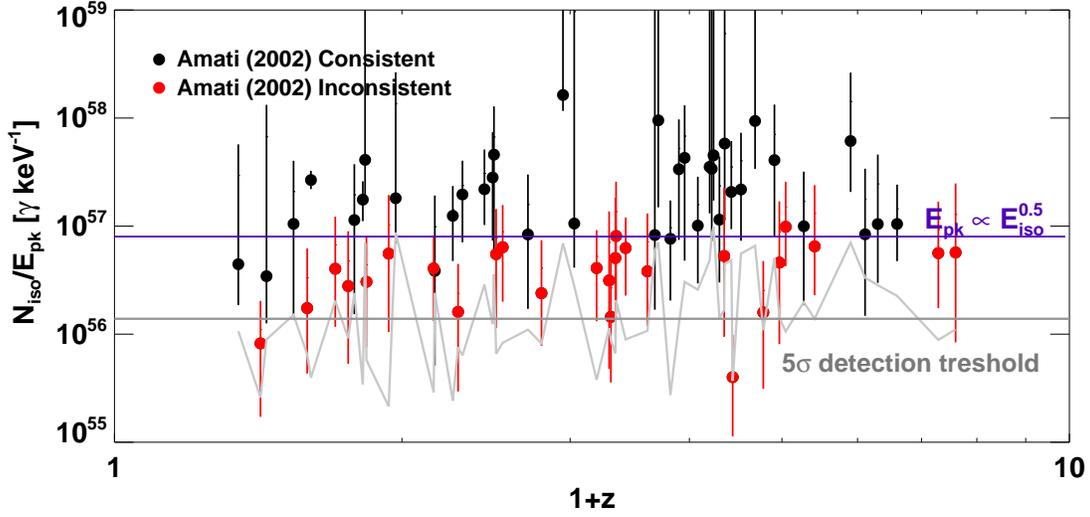}}
\caption{\small $E_{\rm pk}$--$E_{\rm iso}$ truncation and the detector threshold. We 
can separate events consistent with the \citet{amati02} relation (black circles)
from those inconsistent (red circles) by a line of constant $E^{0.5}_{\rm iso}/E_{\rm pk}$, which
is approximately a line of constant $N_{\rm iso}/E_{\rm pk}$.  The detection
threshold scales as $N_{\rm iso}/\sqrt{T_{90}}$ (Section \ref{sec:thresh}).  For the units in the plot,
because of convolution with the
burst $T_{90}$ and $E_{\rm pk}$, the threshold is blurred and the observed instance is a jagged line.
We also plot the median threshold.
Hypothetical events below the purple line (which divides the red and black events and also touches the detection
threshold at various points) are $\gtrsim 65$\% more likely to be absent from pre-{\it Swift}~catalogs.}
\label{fig:amati_thresh}
\end{figure}

\begin{figure}
\centerline{\includegraphics[width=6.0in]{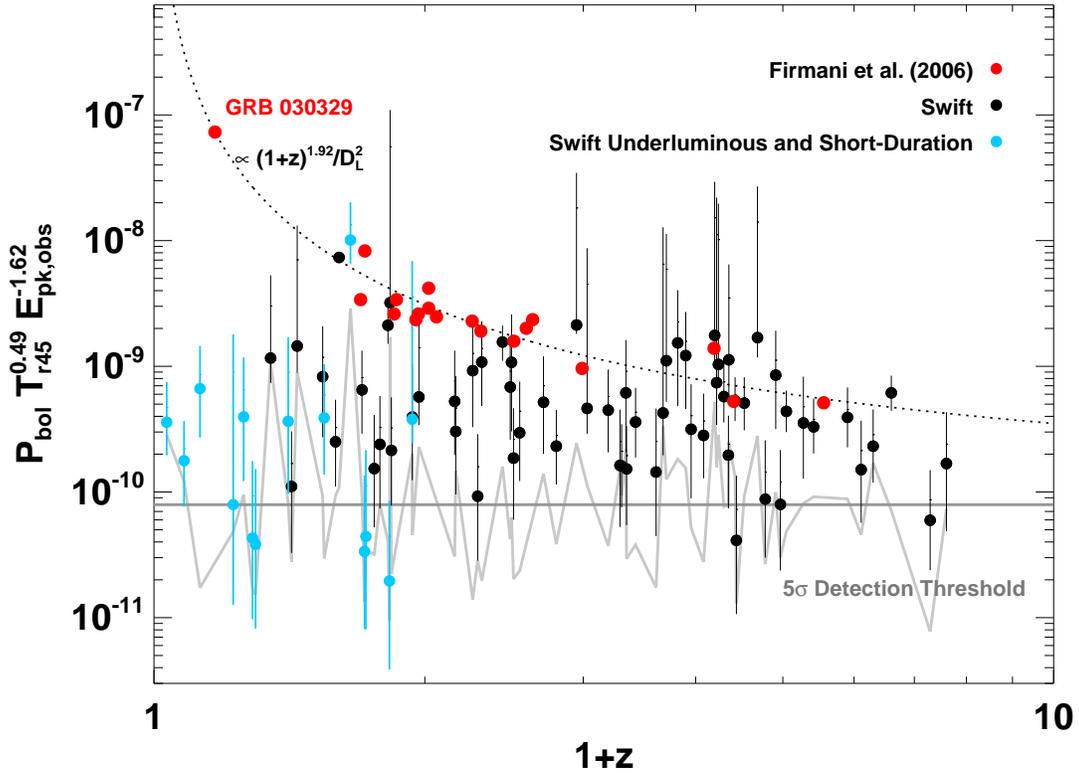}}
\caption{\small The observer frame quantities found to correlate tightly with redshift by \citet{firmani06} do
not correlate tightly in the {\it Swift}~sample.  Both samples follow the same upper limit envelope.  However,
the {\it Swift}~events extend to lower flux and stop at the detector threshold.  Here, $P_{\rm bol}$ is the peak energy flux.  The {\it Swift}~threshold
values (actual and median)
are plotted, as is the best-fit line to the \citet{firmani06} sample.  Error bars are not plotted for the (red) events considered in \citet{firmani06}.}
\label{fig:firmani_thresh}
\end{figure}

There are 3 strong and independent reasons to believe that the $E_{\rm pk}$-$E_{\rm iso}$ relation
for both {\it Swift}~and pre-{\it Swift}~GRBs is an artifact of the detection
threshold.  First, a large fraction of the {\it Swift}~GRB sample exhibits
hard and underluminous spectra which are inconsistent with the pre-{\it Swift}~relations,
in agreement with indications from {\it BATSE}~GRBs without redshifts.  Second,
the {\it Swift}~GRB sample yields a powerlaw $E_{\rm pk}$-$E_{\rm iso}$ relation
which is inconsistent at the $>5\sigma$ level with the pre-{\it Swift} relations,
and with an intrinsic scatter at least a factor of 2 larger.  Third, a dividing
line between the pre-{\it Swift}~and {\it Swift}~samples can be plotted (Figure \ref{fig:amati_thresh}) using only
the detector threshold, and the $E_{\rm pk}$-$E_{\rm iso}$ correlation significance can be shown to decrease
dramatically if we correct for partial correlation with the probable shape of the
threshold.  

These faults appear to be shared by several other correlations
among high-energy parameters reported in the literature (Section \ref{sec:other}).  
It is likely that these contain
largely redundant information which reduces to the shape of the detector threshold,
at least for {\it Swift}~BAT events.  
This insight also helps to explain why short-duration and underluminous
events at low$-z$ appear to fall away from the relations (Section \ref{sec:lowz}).

We stress that even if the relations contain information actually
related to the physical properties of GRBs, the wide dispersion in the relations makes
them useless as cosmology probes \citep[see, also,][]{fb05}.

Nonetheless, it is still likely that the relations encapsulate
important information about the intrinsic distribution of GRBs
with luminosity.  Extracting this information requires that we account accurately for the detector threshold.
Turning this around, at the low energy end, the $E_{\rm pk}$-$E_{\rm iso}$ and other relations may be useful proxies for the
detector threshold and other complicated biases (e.g., those associated with source
localization, optical detection and redshift determination; Section \ref{sec:lowz}).  Accounting for these biases
may be the most fruitful path toward uncovering true source frame relations in GRBs.
This is clearly a critical step toward realizing the potential of {\it Swift}~and future
missions such as {\it GLAST}~and {\it EXIST}.

\acknowledgments
N.~R.~B gratefully acknowledges support from a Townes Fellowship at 
U.~C. Berkeley Space Sciences Laboratory and partial support
from J.~S.~B. and A. Filippenko.  
D.~K. acknowledges financial support through the NSF Astronomy $\&$ Astrophysics Postdoctoral Fellowships under award AST-0502502.
We thank G. Jernigan for useful conversations.  We thank an anonymous
referee for useful criticisms.

\vfill
\eject

\begin{center}
\begin{scriptsize}
\begin{longtable}{l@{}c@{}c@{}cccccc}
\caption[]{BAT Trigger Statistics and Durations} \label{tab:time_table} \\

\hline \hline \\[-2ex]
   GRB & Trigger Time & Burst Region & $T_{90}$ & $T_{50}$ & $T_{r45}$ & Rate$_p/$Cts & $dt_{S/N}$ & $S/N$ \\[0.5ex]
    & (UTC) & [s] & [s] & [s] & [s] & [$1/s$] & [s] & \\[0.5ex] \hline
   \\[-1.8ex]
\endfirsthead

\multicolumn{8}{l}{{\tablename} \thetable{} -- Continued} \\[0.5ex]
  \hline \hline \\[-2ex]
   GRB & Trigger Time & Burst Region & $T_{90}$ & $T_{50}$ & $T_{r45}$ & Rate$_p/$Cts & $dt_{S/N}$ & $S/N$ \\[0.5ex] \hline
  \\[-1.8ex]
\endhead

  \\[-1.8ex] \hline
  \multicolumn{8}{l}{{Continued on Next Page\ldots}} \\[0.5ex]
\endfoot

  \\[-1.8ex] \hline \hline
\endlastfoot
041220 & 22:58:26.00 & $-0.640\rightarrow 3.960$ & $3.5\pm0.2$ & $1.7\pm0.2$ & $1.04\pm0.08$ & $0.6\pm0.1$ & 3.20 & 23.5\\
041223 & 14:06:18.00 & $-10.600\rightarrow 131.000$ & $108.0\pm0.5$ & $29.4\pm0.6$ & $19.2\pm0.4$ & $0.043\pm0.002$ & 60.00 & 177.1\\
041224 & 20:20:57.00 & $-114.610\rightarrow 61.970$ & $141\pm2$ & $42\pm3$ & $28\pm1$ & $0.025\pm0.002$ & 50.22 & 54.9\\
041226 & 20:34:19.00 & $-1.650\rightarrow 8.350$ & $8.1\pm1.0$ & $4.2\pm0.9$ & $2.0\pm0.3$ & $0.3\pm0.1$ & 6.90 & 9.2\\
041228 & 10:49:13.00 & $-2.465\rightarrow 69.315$ & $56\pm6$ & $20\pm1$ & $16.3\pm0.8$ & $0.037\pm0.005$ & 37.00 & 42.2\\
050117 & 12:52:36.00 & $0.900\rightarrow 213.860$ & $169\pm1$ & $84\pm1$ & $48\pm1$ & $0.019\pm0.001$ & 174.24 & 78.4\\
050124 & 11:30:03.00 & $-2.200\rightarrow 2.680$ & $3.4\pm0.3$ & $2.00\pm0.08$ & $1.08\pm0.06$ & $0.57\pm0.08$ & 3.56 & 36.9\\
050126 & 12:00:53.00 & $-1.905\rightarrow 35.505$ & $29\pm2$ & $15\pm1$ & $8.1\pm0.8$ & $0.07\pm0.02$ & 26.39 & 16.2\\
\end{longtable}
{Notes: We quote durations and time statistics for extended trigger windows around three
short bursts (GRBs 050724, 051227, and 061006; Section \ref{sec:caveats}).}
\end{scriptsize}
\end{center}

\vfill
\eject

\begin{center}
\begin{scriptsize}
\begin{longtable}{@{}l@{}c@{}c@{}c@{}c@{}c@{}c@{}c@{}c@{}c}
\caption[]{BAT Spectral Fits} \label{tab:spec_table} \\

\hline \hline \\[-2ex]
   GRB & -$\alpha$ & -$\beta$ & $E^{\rm freq.}_{\rm pk,obs}$ & $S_{15-350}$ & $\chi^2/\nu^{({\rm model})}$ & $E_{\rm pk,obs}$ & 
   $N_{\rm iso}$ or $n_{\rm bol}$ & $E_{\rm iso}$ or $S_{\rm bol}$ & $z^{({\rm ref})}$ \\[0.5ex]
    &  & & [keV] & [$10^{-6}{{\rm erg} \over {\rm cm}^2}$] &  & [keV] & [$10^{59}\gamma$ or $10^2{\gamma \over {\rm cm}^2}$] & 
    [$10^{52}$erg or $10^{-6}{{\rm erg} \over {\rm cm}^2}$] & \\[0.5ex] \hline
   \\[-1.8ex]
\endfirsthead

\multicolumn{10}{l}{{\tablename} \thetable{} -- Continued} \\[0.5ex]
  \hline \hline \\[-2ex]
   GRB & -$\alpha$ & -$\beta$ & $E^{\rm freq.}_{\rm pk,obs}$ & $S_{15-350~{\rm keV}}$ & $\chi^2/\nu^{({\rm model})}$ & $E_{\rm pk,obs}$ & 
   $N_{\rm iso}$ or $n_{\rm bol}$ & $E_{\rm iso}$ or $S_{\rm bol}$ & $z^{({\rm ref})}$ \\[0.5ex] \hline
  \\[-1.8ex]
\endhead

  \\[-1.8ex] \hline
  \multicolumn{10}{l}{{Continued on Next Page\ldots}} \\[0.5ex]
\endfoot

  \\[-1.8ex] \hline \hline
\endlastfoot
041220 & $1.6\pm0.1$ & ... & $>$87 & $0.57\pm0.04$ & $54.6/55^{1}$ & $111^{+398}_{-38}$ & $0.17^{+0.13}_{-0.04}$ & $0.6^{+0.6}_{-0.1}$ & ...\\
041223 & $0.9\pm0.1$ & ... & $400^{+400}_{-100}$ & $41\pm2$ & $30.9/54^{2}$ & $360^{+340}_{-100}$ & $5.4\pm0.7$ & $120^{+30}_{-40}$ & ...\\
041224 & $0.7\pm0.3$ & ... & $74^{+10}_{-6}$ & $9.5\pm0.5$ & $26.1/54^{2}$ & $76\pm9$ & $3.0^{+0.9}_{-0.6}$ & $12^{+2}_{-1}$ & ...\\
041226 & $1.4\pm0.3$ & ... & $>$70 & $0.27^{+0.06}_{-0.08}$ & $68.0/55^{1}$ & $162^{+651}_{-94}$ & $0.06^{+0.05}_{-0.02}$ & $0.3^{+0.5}_{-0.1}$ & ...\\
041228 & $1.54\pm0.08$ & ... & $>$152 & $6.3^{+0.3}_{-0.6}$ & $43.8/55^{1}$ & $240^{+630}_{-110}$ & $2.2^{+0.9}_{-0.5}$ & $9^{+6}_{-3}$ & ...\\
050117 & $1.1\pm0.2$ & ... & $130^{+70}_{-30}$ & $13.1\pm0.9$ & $45.0/54^{2}$ & $135^{+73}_{-26}$ & $3.6^{+0.9}_{-0.7}$ & $16^{+4}_{-2}$ & ...\\
050124 & $0.5\pm0.4$ & ... & $90^{+20}_{-10}$ & $1.43^{+0.12}_{-0.10}$ & $49.3/54^{2}$ & $101^{+35}_{-15}$ & $0.33^{+0.14}_{-0.05}$ & $1.8^{+0.4}_{-0.2}$ & ...\\
050126 & $1.3\pm0.2$ & ... & $>$85 & $1.9^{+0.2}_{-0.8}$ & $59.2/55^{1}$ & $169^{+496}_{-63}$ & $0.6^{+0.5}_{-0.2}$ & $0.8^{+1.0}_{-0.2}$ & $1.29^{1}$\\
\end{longtable}
{Notes: The model in column 6 refers to a powerlaw (1), a powerlaw times exponential (2), a GRBM (3).  
If a redshift is known, it is given in the last table column and we present isotropic equivalent energy 
and photon fluences, $E_{\rm iso}$ and $N_{\rm iso}$, respectively, in columns 8 and 9.  Otherwise, we report approximate
bolometric fluences measured in the observer frame $1-10^4$ keV band in those columns.
$^*$We quote spectral fits for extended trigger windows around three
short bursts (GRBs 050724, 051227, and 061006; Section \ref{sec:caveats}).
$^{\rm \dag}$These events have $E_{\rm pk,obs}$ measurements
from Konus-Wind \citep[e.g.,][]{gol06}:
050326 $201\pm24$ keV,
050401 $132\pm16$ keV,
050525 $84\pm1$ keV,
050603 $349\pm28$ keV,
050713A $312\pm50$ keV,
050717 $1890^{+1600}_{-760}$ keV,
051008 $870^{+180}_{-140}$ keV,
051221A $402^{+93}_{-72}$ keV,
060105 $549\pm34$ keV,
060117 $89\pm5$ keV,
060313 $920^{+310}_{-180}$ keV,
060418 $230\pm23$ keV,
060510A $184^{+36}_{-24}$ keV,
060614 $302^{+214}_{-85}$ keV,
060813 $192\pm20$ keV,
060814 $257^{+122}_{-58}$ keV,
060904A $163\pm31$ keV,
061006 $660^{+230}_{-140}$ keV,
061007 $399\pm19$ keV,
061021 $780^{+550}_{-240}$ keV,
061121 $606^{+90}_{-72}$ keV,
061201 $870^{+460}_{-280}$ keV,
061222A $283^{+59}_{-42}$ keV,
070220 $315^{+70}_{-115}$ keV,
070328 $690^{+170}_{-120}$ keV,
070420 $147^{+29}_{-19}$ keV,
and 070508 $188\pm8$ keV.
$^{\rm \ddag}$These events have $E_{\rm pk,obs}$ measurements
from Suzaku \citep[e.g.,][]{hong07}:
GRB051008 $1540^{+560}_{-1420}$ keV,
GRB060813 $248\pm20$ keV,
GRB060814 $310\pm159$ keV,
GRB061006 $970^{+390}_{-210}$ keV,
GRB061007 $561\pm29$ keV,
GRB070328 $960^{+1010}_{-450}$ keV,
and GRB070508 $233\pm12$ keV.}
{Redshift References:
$^{1}$\citet{zref1},
$^{2}$\citet{zref2},
$^{3}$\citet{zref3},
$^{4}$\citet{zref3},
$^{5}$\citet{zref5},
$^{6}$\citet{zref6},
$^{7}$\citet{zref7},
$^{8}$\citet{zref8},
$^{9}$\citet{zref9},
$^{10}$\citet{zref10},
$^{11}$\citet{zref11},
$^{12}$\citet{zref12},
$^{13}$\citet{zref13},
$^{14}$\citet{zref14},
$^{15}$\citet{zref15},
$^{16}$\citet{zref16},
$^{17}$\citet{zref17},
$^{18}$\citet{zref18},
$^{19}$\citet{zref19},
$^{20}$\citet{zref20},
$^{21}$\citet{zref21},
$^{22}$\citet{zref22},
$^{23}$\citet{zref5},
$^{24}$\citet{zref24},
$^{25}$\citet{zref25},
$^{26}$\citet{zref26},
$^{27}$\citet{zref27},
$^{28}$\citet{zref28},
$^{29}$\citet{zref29},
$^{30}$\citet{zref30},
$^{31}$\citet{zref31},
$^{32}$\citet{zref32},
$^{33}$\citet{zref33},
$^{34}$\citet{zref34},
$^{35}$\citet{zref35},
$^{36}$\citet{zref36},
$^{37}$\citet{zref37},
$^{38}$\citet{zref38},
$^{39}$\citet{zref39},
$^{40}$\citet{zref40},
$^{41}$\citet{zref41},
$^{42}$\citet{zref42},
$^{43}$\citet{zref43},
$^{44}$\citet{zref44},
$^{45}$\citet{zref45},
$^{46}$\citet{zref5},
$^{47}$\citet{zref47},
$^{48}$\citet{zref48},
$^{49}$\citet{zref49},
$^{50}$\citet{zref50},
$^{51}$\citet{zref5},
$^{52}$\citet{zref52},
$^{53}$\citet{zref5},
$^{54}$\citet{zref54},
$^{55}$\citet{zref55},
$^{56}$\citet{zref5},
$^{57}$\citet{zref57},
$^{58}$\citet{zref58},
$^{59}$\citet{zref59},
$^{60}$\citet{zref60},
$^{61}$\citet{zref61},
$^{62}$\citet{zref62},
$^{63}$\citet{zref63},
$^{64}$\citet{zref64},
$^{65}$\citet{zref65},
$^{66}$\citet{zref66},
$^{67}$\citet{zref67},
$^{68}$\citet{zref67},
$^{69}$\citet{zref69},
$^{70}$\citet{zref70},
$^{71}$\citet{zref71},
$^{72}$\citet{zref72},
$^{73}$\citet{zref73},
$^{74}$\citet{zref74},
$^{75}$\citet{zref75},
$^{76}$\citet{zref76},
and $^{77}$\citet{zref77}.
}
\end{scriptsize}
\end{center}

\vfill
\eject

\end{document}